\documentstyle[12pt]{article}
\author{Yu.N.Bespalov}
\title{\mbox{Crossed modules}\\
       \mbox{and}\\
       \mbox{quantum groups}\\
       \mbox{in}\\
       \mbox{braided categories $\;$ I.}}
\data{May 6, 1994}

\def\ltimes{\mathrel{\,\hbox{\vrule height 4.5pt}\!\times}}
\def\newfigure{\begin{picture}(0,0)\end{picture}\newpage}
\def\interskip{\bigskip}
\newtheorem{theorem}{\bf Theorem}[section]
\newtheorem{proposition}{\bf Propositon}[section]
\newtheorem{lemma}{\bf Lemma}[section]
\newtheorem{definition}{\bf Definition}[section]
\newtheorem{remark}{\bf Remark}[section]
\newtheorem{example}{\bf Example}[section]
\newenvironment{proof}{\par\noindent{\bf Proof.}}{$\quad
\hbox{\vrule height 7pt width 0.5pt depth 1pt}\!
      \lower -2pt\hbox{${\vbox {\hrule width 8pt height 0.5pt depth 0pt}\atop
      \vbox {\hrule width 8pt height 0.5pt depth 0pt}}$}
      \!\hbox{\vrule height 7pt width 0.5pt depth 1pt}$\interskip}
\def\krr{\kern -.16667em}%
\def\kr{}%
\def\krrr{\kern -.3\unitlength}%
\newlength{\textwd}%
\def\hhstep{\kr\kr
\kern -.5\unitlength}
\def\hstep{\kr\kr
\kern .5\unitlength}
\def\step{\kr\kr
\kern \unitlength}
\def\Step{\kr\kr
\kern 2\unitlength}
\def\vvbox#1{{\offinterlineskip\vcenter{%
\def\coev{\kr
\begin{picture}(2,2)\put(1,0){\oval(2,2)[t]}\end{picture}}
\def\ev{\kr
\begin{picture}(2,2)\put(1,2){\oval(2,2)[b]}\end{picture}}
\def\hcoev{\kr
\begin{picture}(1,2)\put(.5,0){\oval(1,1)[t]}\end{picture}}
\def\hev{\kr
\begin{picture}(1,2)\put(.5,2){\oval(1,1)[b]}\end{picture}}
\def\COEV{\kr
\begin{picture}(2,2)\put(3,0){\oval(6,6)[t]}\end{picture}}
\def\EV{\kr
\begin{picture}(2,2)\put(3,2){\oval(6,6)[b]}\end{picture}}
\def\unit{\kr
\begin{picture}(0,2)
\put(0,0){\line(0,1){1}}\put(0,1.2){\circle{0.4}}
\end{picture}}
\def\counit{\kr
\begin{picture}(0,2)
\put(0,1){\line(0,1){1}}\put(0,.8){\circle{0.4}}
\end{picture}}
\def\Q##1{\kr
\begin{picture}(0,2)
\put(0,0){\line(0,1){0.4}}\put(0,1){\circle{1.2}}
\put(-0.6,0.4){\makebox(1.2,1.2)[cc]{$\scriptstyle ##1$}}
\end{picture}}
\def\O##1{\kr
\begin{picture}(0,2)
\put(0,0){\line(0,1){0.4}}\put(0,1.6){\line(0,1){0.4}}\put(0,1){\circle{1.2}}
\put(-0.6,0.4){\makebox(1.2,1.2)[cc]{$\scriptstyle ##1$}}
\end{picture}}
\def\S{\O{S}}                 \def\SS{\O{S^{-1}}}
\def\tS{\O{\widetilde S}}     \def\tSS{\O{\widetilde S^{-1}}}
\def\x{\kr
\begin{picture}(2,2)
\put(0,2){\line(1,-1){2}}\put(0,0){\line(1,1){.7}}\put(2,2){\line(-1,-1){.7}}
\end{picture}}
\def\xx{\kr
\begin{picture}(2,2)
\put(0,2){\line(1,-1){.7}}\put(0,0){\line(1,1){2}}\put(2,0){\line(-1,1){.7}}
\end{picture}}
\def\hx{\kr
\begin{picture}(1,2)
\put(0,2){\line(1,-2){1}}\put(0,0){\line(1,2){.35}}\put(1,2){\line(-1,-2){.35}}
\end{picture}}
\def\hxx{\kr
\begin{picture}(1,2)
\put(0,2){\line(1,-2){.35}}\put(0,0){\line(1,2){1}}\put(1,0){\line(-1,2){.35}}
\end{picture}}
\def\d{\kr
\begin{picture}(1,2)\put(0,2){\line(1,-2){1}}\end{picture}}
\def\dd{\kr
\begin{picture}(1,2)\put(0,0){\line(1,2){1}}\end{picture}}
\def\hd{\kr
\begin{picture}(1,2)
\put(0,2){\line(1,-2){.5}}
\put(.5,1){\line(0,-1){1}}
\end{picture}}
\def\hdd{\kr
\begin{picture}(1,2)
\put(1,2){\line(-1,-2){.5}}
\put(0,1){\line(0,-1){1}}
\end{picture}}
\def\ld{\kr
\begin{picture}(1,2)
\put(1,0){\oval(2,2)[lt]}\put(1,0){\line(0,1)2}
\end{picture}}
\def\Ld{\kr
\begin{picture}(2,2)
\put(2,0){\oval(4,2)[lt]}\put(2,0){\line(0,1)2}
\end{picture}}
\def\cd{\kr
\begin{picture}(2,2)
\put(1,0){\oval(2,2)[ct]}\put(1,1){\line(0,1)1}
\end{picture}}
\def\hdcd{\kr
\begin{picture}(1,2)
\put(0,2){\line(1,-2){.5}}
\put(.5,0){\oval(1,1)[ct]}\put(.5,.5){\line(0,1){.5}}
\end{picture}}
\def\hddcd{\kr
\begin{picture}(1,2)
\put(1,2){\line(-1,-2){.5}}
\put(.5,0){\oval(1,1)[ct]}\put(.5,.5){\line(0,1){.5}}
\end{picture}}
\def\hcd{\kr
\begin{picture}(1,2)
\put(.5,0){\oval(1,1)[ct]}\put(.5,.5){\line(0,1){1.5}}
\end{picture}}
\def\Cd{\kr
\begin{picture}(4,2)
\put(2,0){\oval(4,2)[ct]}\put(2,1){\line(0,1)1}
\end{picture}}
\def\rd{\kr
\begin{picture}(1,2)
\put(0,0){\oval(2,2)[rt]}\put(0,0){\line(0,1)2}
\end{picture}}
\def\Rd{\kr
\begin{picture}(2,2)
\put(0,0){\oval(4,2)[rt]}\put(0,0){\line(0,1)2}
\end{picture}}
\def\lu{\kr
\begin{picture}(1,2)
\put(1,2){\oval(2,2)[lb]}\put(1,0){\line(0,1)2}
\end{picture}}
\def\Lu{\kr
\begin{picture}(2,2)
\put(2,2){\oval(4,2)[lb]}\put(2,0){\line(0,1)2}
\end{picture}}
\def\cu{\kr
\begin{picture}(2,2)
\put(1,2){\oval(2,2)[cb]}\put(1,0){\line(0,1)1}
\end{picture}}
\def\hcu{\kr
\begin{picture}(1,2)
\put(.5,2){\oval(1,1)[cb]}\put(.5,0){\line(0,1){1.5}}
\end{picture}}
\def\Cu{\kr
\begin{picture}(4,2)
\put(2,2){\oval(4,2)[cb]}\put(1,0){\line(0,1)1}
\end{picture}}
\def\ru{\kr
\begin{picture}(1,2)
\put(0,2){\oval(2,2)[rb]}\put(0,0){\line(0,1)2}
\end{picture}}
\def\Ru{\kr
\begin{picture}(2,2)
\put(0,2){\oval(4,2)[rb]}\put(0,0){\line(0,1)2}
\end{picture}}
\def\k{\kr
\begin{picture}(1,2)
\put(0,2){\oval(2,1)[rb]}
\put(0,0){\oval(2,1)[rt]}
\put(0,0){\line(0,1)2}
\end{picture}}
\def\ro##1{\kr
\begin{picture}(2,2)
\put(.4,0){\oval(.8,.8)[lt]}\put(1.6,0){\oval(.8,.8)[rt]}
\put(1,0.4){\circle{1.2}}
\put(0.4,-0.2){\makebox(1.2,1.2)[cc]{$\scriptstyle ##1$}}%
\end{picture}}
\def\Ro##1{\kr
\begin{picture}(4,2)
\put(1.4,0){\oval(2.8,1.2)[lt]}\put(2.6,0){\oval(2.8,1.2)[rt]}
\put(2,.6){\circle{1.2}}
\put(1.4,0){\makebox(1.2,1.2)[cc]{$\scriptstyle ##1$}}%
\end{picture}}
\def\r{\ro{\cal R}}           \def\rr{\ro{{\cal R}^{-1}}}
\def\ra{\ro{{\cal R}_A}}        \def\rra{\ro{{\cal R}^{-1}_A}}
\def\rb{\ro{{\cal R}_B}}        \def\rrb{\ro{{\cal R}^{-1}_B}}
\def\rh{\ro{{\cal R}_H}}
\def\R{\Ro{\cal R}}           \def\RR{\Ro{{\cal R}^{-1}}}
\def\Ra{\Ro{{\cal R}_A}}        \def\RRa{\Ro{{\cal R}^{-1}_A}}
\def\Rb{\Ro{{\cal R}_B}}        \def\RRb{\Ro{{\cal R}^{-1}_B}}
\def\Rh{\Ro{{\cal R}_H}}
\def\id{\kr
\begin{picture}(0,2)\put(0,0){\line(0,1)2}\end{picture}}
\def\obj##1{\settowidth{\textwd}{$##1$}%
\raise .2\unitlength\hbox{\kern -.5\textwd $##1$ \kern -.5\textwd \krrr}}
\def\Obj##1{\settowidth{\textwd}{$##1$}%
\raise 1.1\unitlength\hbox{\kern -1\textwd $##1$}}
\def\hhbox##1{\hbox{%
\kern -4.45\unitlength
\def\coev{\kr
\begin{picture}(1,1)\put(.5,0){\oval(1,1)[t]}\end{picture}}
\def\ev{\kr
\begin{picture}(1,1)\put(.5,1){\oval(1,1)[b]}\end{picture}}
\def\ld{\kr
\begin{picture}(1,1)
\put(1,0){\oval(2,2)[lt]}\put(1,0){\line(0,1)1}
\end{picture}}
\def\Ld{\kr
\begin{picture}(2,1)
\put(2,0){\oval(4,2)[lt]}\put(2,0){\line(0,1)1}
\end{picture}}
\def\rd{\kr
\begin{picture}(1,1)
\put(0,0){\oval(2,2)[rt]}\put(0,0){\line(0,1)1}
\end{picture}}
\def\Rd{\kr
\begin{picture}(2,1)
\put(0,0){\oval(4,2)[rt]}\put(0,0){\line(0,1)1}
\end{picture}}
\def\cd{\kr
\begin{picture}(1,1)
\put(.5,0){\oval(1,1)[ct]}\put(.5,.5){\line(0,1){.5}}
\end{picture}}
\def\lu{\kr
\begin{picture}(1,1)
\put(1,1){\oval(2,2)[lb]}\put(1,0){\line(0,1)1}
\end{picture}}
\def\Lu{\kr
\begin{picture}(2,1)
\put(2,1){\oval(4,2)[lb]}\put(2,0){\line(0,1)1}
\end{picture}}
\def\cu{\kr
\begin{picture}(1,1)
\put(.5,1){\oval(1,1)[cb]}\put(.5,0){\line(0,1){.5}}
\end{picture}}
\def\ru{\kr
\begin{picture}(1,1)
\put(0,1){\oval(2,2)[rb]}\put(0,0){\line(0,1)1}
\end{picture}}
\def\Ru{\kr
\begin{picture}(2,1)
\put(0,1){\oval(4,2)[rb]}\put(0,0){\line(0,1)1}
\end{picture}}
\def\hru{\kr
\begin{picture}(.5,1)
\put(0,1){\oval(1,1)[rb]}\put(0,0){\line(0,1)1}
\end{picture}}
\def\hrd{\kr
\begin{picture}(.5,1)
\put(0,0){\oval(1,1)[rt]}\put(0,0){\line(0,1)1}
\end{picture}}
\def\id{\kr
\begin{picture}(0,1)\put(0,0){\line(0,1)1}\end{picture}}
\def\d{\kr
\begin{picture}(.5,1)\put(0,1){\line(1,-2){0.5}}\end{picture}}
\def\dd{\kr
\begin{picture}(.5,1)\put(0,0){\line(1,2){0.5}}\end{picture}}
##1}}
#1}\normalbaselines}}
\def\object#1{\settowidth{\textwd}{$#1$}%
                        \hbox{%
                        \kern -.5\textwd $#1$ \kern -.5\textwd}}
\def\map#1#2#3{\vcenter{\hbox{$#2\;$}}
                     \vcenter{\settowidth{\textwd}{$#1$}
	                      \hbox{\kern -.5\textwd $#1$ \kern -.5\textwd}
			      \hbox{\begin{picture}(0,2)
                                          \put(0,2){\vector(0,-1)2}
                                    \end{picture}}
                              \settowidth{\textwd}{$#3$}
	                      \hbox{\kern -.5\textwd $#3$ \kern -.5\textwd}}}

\begin{document}
\maketitle
\begin{abstract}
Let $H$ be a Hopf algebra in braided category $\cal C$.
Crossed modules over $H$ are objects with both module and comodule structures
satisfying some comatibility condition.
Category ${\cal C}^H_H$ of crossed modules is braided and is concrete
realization of general categorical construction.
For quantum braided group $(H,{\cal R})$ corresponding braided category
${\cal C}^{\cal R}_H$ of modules is identifyed with full subcategory in
${\cal C}_H^H$.
Connection with crossproducts is discussed.
Correct cross product in the class of quantum braided groups is built.
Radford's--Majid's theorem gives equivalent condition for usual Hopf algebra
to be crossproduct.
Braided variant and analog of this theorem for quantum braided qroups
are obtained.
\end{abstract}

\section{Introduction}  

Crossed modules over finite group $G$ was considered in \cite{Whitehead1}.
Generalization for arbitrary Hopf algebra $H$ belongs to Yetter \cite{Yetter1}.
Crossed module is vector space with both module and comodule structures
over $H$ satisfying some comatibility condition.
Category ${}_H^H{\cal M}$ of crossed modules is braided and
if $H$ is finite dimensional coinside with category of modules over Drinfeld's
quantum double ${\cal D}(H)$ \cite{Drinfeld1}.
One can obtain ${}_H^H{\cal M}$ also as 'center' or 'inner double' of monoidal
category ${}_H{\cal M}$ of left modules.
Crossed modules appear in different contexts related with Hopf algebras
and quantum groups.
For example, subalgebra of left or right invariant forms in bicovariant
differential algebra over Hopf algebra has a crossed module structure.

In this paper we introduce and study a category ${\cal C}_H^H$
of crossed modules over Hopf algebra $H$
liveing in arbitrary braided monoidal category $\cal C$
({\em braided Hopf algebra}
\cite{Lyubashenko1}-\cite{Lyubashenko4},\cite{Majid6}-\cite{Majid10}).
This category is also braded and most of results for usual Hopf algebras
keep in this situation.
But if even there exist dual braded Hopf algebra $H^\vee$ category
${\cal C}_H^H$, generaly speaking, can not be realized as a category of
modules over something like quantum double.
So the notion of crossed module is more general then of quantum double.

Majid \cite{Majid6}-\cite{Majid10} define quantum braided group as a pair:
braided Hopf algebra $H$ with 'element' $\cal R$, satisfying some axioms
which are not trivial generalization of ones for usual quantum group and
appropriate class $\cal O$ of modules over $H$.
We will consider greatest such class ${\cal C}_H^{\cal R}$
and use the same notation for category with these objects.
Category ${\cal C}^{\cal R}_H$ is braided and can be identifyed with
full subcategory in ${\cal C}_H^H$.
In \cite{Majid6}-\cite{Majid10} left modules over quantum braided group are
mainly used. In this case there also exists imbedding of braided categories
${}^{\cal R}_H{\cal C}\hookrightarrow{}_H^H{\cal C}$.
But formulas for right modules are simple.
Identification modules over quantum braided groups with crossed modules has
both strict and heuristical applications for example to quantum group
cross products.

Let $A$ be a Hopf algebra in braided category $\cal C$ and $B$ a Hopf algebra
in category ${\cal C}_A^A$ of crossed modules. Then like in unbraided case
\cite{Majid10} tensor product $A\otimes B$ can be equiped with natural
cross product Hopf algebra structure $A\ltimes B$.
Radford's--Majid's theorem \cite{Majid10} gives equivalent condition for
usual Hopf algebra to be cross product.
Braided variant of this theorem is also true if we suppose that idempotents
in category $\cal C$ are split. The last condition is not essential because
any braided category can be extended up to one with split idempotents.

Similarly let $(A,{\cal R}_A)$ be a quantum group in category $\cal C$
and $(B,{\cal R}_B)$ quantum group in ${\cal C}_H^{\cal R}$.
Then cross product Hopf algebra $A\ltimes B$ is also quantum group with
${\cal R}_{A\ltimes B}$ built from ${\cal R}_A$ and ${\cal R}_B$.
This construction generalize Majid's 'bosonization' which was done in
\cite{Majid6} with restrictions like cocommutativity or trivial
braiding.
Equivalent condition for quantum braided group to be cross product
are obtained.

Results related with this paper in the case of usual Hopf algebras
were considered in \cite{Bespalov1}.

\interskip
In section 2 necessary notions and results connected with braided categories
and braided Hopf algebras are done.

In section 3 category ${\cal C}^H_H$ of crossed modules over Hopf algebra $H$
in braided category $\cal C$ is introduced and studied.

Section 4 is devoted to cross products of braded Hopf algebras and
braided variant of Radford's--Majid's theorem.

Subjects related with quantum braided groups will be considered in second
part of this paper.

\section{Braided categories and braided Hopf algebras.} 

In this section we will give short summary of results connected with
braided categories and Hopf algebras in them.
We follow basically \cite{Majid8},\cite{Majid10}.
See also \cite{Freyd1}-\cite{Lyubashenko4}.

\subsection{Braided categories.}

The conception of the braided category generalise the notion of
supersymmetry and group covariance.
This structure appier in connection with quantum groups, in Galois theory,
in constructions of knot invariants, in quantum field theory, etc.

\begin{definition}
{\em Monoidal category}
(${\cal C}$, $\otimes$, $\underline 1$, $\Phi$)
is the category ${\cal C}$ equipped with
\begin{itemize}
\item
the functor  $\quad \otimes : \cal C \times \cal C \rightarrow \cal C \quad $
{\em (tensor product)};
\item
the object $\underline 1$ with functorial isomorphisms
 $\quad \underline 1 \otimes X\simeq X\simeq X\otimes \underline 1 \quad$
{\em (unit object)};
\item
{associativity functorial isomorphisms}
\begin{displaymath}
\Phi = \Phi _{X,Y,Z}: X\otimes (Y\otimes Z) \buildrel \sim \over \rightarrow
   (X\otimes Y)\otimes Z
\end{displaymath}
obeying the pentagon coherence identity
\begin{displaymath}
\matrix{
	  & (X\otimes Y)\otimes (Z\otimes T)  && \cr
          {}^{\Phi_{X,Y,Z\otimes T}}\nearrow &&
          \searrow^{\Phi_{X\otimes Y,Z,T}} & \cr
	  &&&\cr
	  X\otimes (Y\otimes (Z\otimes T))  &&
	  ((X\otimes Y)\otimes Z)\otimes T  & \cr        &&&\cr
          \hfil{}^{{\rm id}\otimes\Phi_{Y,Z,T}}\downarrow\hfil\hfil &&
          \hfil\hfil\uparrow^{\Phi_{X,Y,Z}\otimes{\rm id}}\hfil & \cr
	  X\otimes ((Y\otimes Z)\otimes T)  &
          \buildrel{\Phi_{X,Y\otimes Z,T}}\over\longrightarrow  &
	  (X\otimes (Y\otimes Z))\otimes T  &\cr                        }
\end{displaymath}
\end{itemize}
\end{definition}

In basic examples associativity isomorphisms $\Phi$ unit object $\underline 1$
and its associated maps are typically the obvious ones, so we will not write
them to explicitely.
However, they should be understood in all formulas.

\begin{definition}
Monoidal category ${\cal C}$ is called {\em braided} if in addition there
exists functorial braiding isomorphisms
\begin{displaymath}
\Psi =\Psi _{X,Y}: X\otimes Y \buildrel \sim \over \rightarrow Y\otimes X
\end{displaymath}
well-behaved under tensor products of objects:
\begin{displaymath}
\Psi_{X\otimes Y,Z}=(\Psi_{X,Z}\otimes{\rm id})({\rm id}\otimes\Psi_{Y,Z}),
\qquad
\Psi_{X,Y\otimes Z}=({\rm id}\otimes\Psi_{X,Z})(\Psi_{X,Y}\otimes{\rm id}).
\end{displaymath}
If we put in $\Phi$ and write these conditions as commutative diagramms
they look like hexagons.
\end{definition}

We will use the notations:

\begin{equation}
\map{X\otimes Y}{{}^{\cal C}\Psi_{X,Y}}{Y\otimes X}\qquad
={}\quad
\matrix{\object{X}\Step\object{Y}\cr
        \vvbox{\hbox{\x}}\cr
	\object{Y}\Step\object{X}}
\qquad\qquad
\map{X\otimes Y}{{}^{\cal C}\Psi_{Y,X}^{-1}}{Y\otimes X}\qquad
={}\quad
\matrix{\object{X}\Step\object{Y}\cr
        \vvbox{\hbox{\xx}}\cr
	\object{Y}\Step\object{X}}
\label{Psi}
\end{equation}

The coherence theorem for braided categories says that if we are given two
composites of the $\Psi$, $\Phi$ etc. and write them as braides according to
(\ref{Psi}) (suppressing the $\Phi$) then the compositions are the same if the
corresponding braides are topologically, the same.

\begin{definition}
(Notations are from \cite{Lyubashenko1}.)
The object ${}^\vee X$ called to be {\em left dual} for object $X$ of monoidal
category ${\cal C}$ if there exist morphisms
\begin{displaymath}
\map{{}^\vee X\otimes X}{{\rm ev}_X}{\underline 1}
\quad ={}\quad{}
\matrix{\object{{}^\vee X}\Step\object{X}\cr
        \vvbox{\hbox{\ev}}}
\qquad\qquad
\map{\underline 1}{{\rm coev}_X}{X\otimes{}^\vee X}
\quad ={}\quad{}
\matrix{\vvbox{\hbox{\coev}}\cr
        \object{X}\Step\object{{}^\vee X}}
\end{displaymath}
obeying
\begin{equation}
\matrix{\object{{}^\vee X}\Step\cr
	\vvbox{\hbox{\id\step\hcoev}
	       \hbox{\hev\step\id}}\cr
	\Step\Step\object{{}^\vee X}}
\quad =\quad
\matrix{\object{{}^\vee X}\cr
        \vvbox{\hbox{\id}\hbox{\id}}\cr
        \object{{}^\vee X}}
\qquad\qquad
\matrix{\Step\Step\object{X}\cr
	\vvbox{\hbox{\hcoev\step\id}
	       \hbox{\id\step\hev}}\cr
	\object{X}\Step}
\quad ={}\quad{}
\matrix{\object{X}\cr
        \vvbox{\hbox{\id}\hbox{\id}}\cr
        \object{X}}
\end{equation}

{\em Right dual object} $X^\vee$ is defined similarly with maps and identities
wich look like mirror reflections of ones for left-dual.
\end{definition}

{\em (Braided) monoidal functor} $F$ is functor between (braided) monoidal
categories together with functorial isomorphisms
\begin{equation}
F(X)\otimes F(Y) \simeq F(X\otimes Y)
\end{equation}
coordinated with associativity isomorphisms (and braiding).

\subsection{Hopf algebras.}

\begin{definition}
\begin{itemize}
\item
{\em Algebra} in monoidal category $\cal C$ is object $A$ equipped with
unit $\eta$ and multiplication $\cdot$
\begin{displaymath}
\map{\underline 1}{\eta}{A}
\quad\!\! ={}\quad\;
\vvbox{\hbox{\unit}}
\qquad\qquad
\map{A\otimes A}{\cdot}{A}
\quad ={}\quad{}
\vvbox{\hbox{\cu}}
\end{displaymath}
obeying
\begin{equation}
\vvbox{\hbox{\unit\Step\id}\hbox{\cu}}
\quad ={}\quad{}
\vvbox{\hbox{\id}\hbox{\id}}
\quad ={}\quad{}
\vvbox{\hbox{\id\Step\unit}\hbox{\cu}}
\qquad\qquad
\vvbox{\hbox{\cu\step\id}
       \hbox{\step\cu}}
\quad ={}\quad{}
\vvbox{\hbox{\id\step\cu}
       \hbox{\cu}}
\label{algebra}
\end{equation}
\item
{\em Coalgebra} is object $C$ equipped with counit $\epsilon$ and
comultiplication $\Delta$
\begin{displaymath}
\map{C}{\epsilon}{\underline 1}
\quad\!\! ={}\quad{}\;
\vvbox{\hbox{\counit}}
\qquad\qquad
\map{C}{\Delta}{C\otimes C}
\quad ={}\quad{}
\vvbox{\hbox{\cd}}
\end{displaymath}
obeying axioms like (\ref{algebra}) turned upside-doun.
\end{itemize}
\end{definition}

\begin{proposition} \cite{Majid8}\cite{Majid10}
Let $A$, $B$ be two algebras in braided category $\cal C$. There is a braided
tensor product algebra $A\underline\otimes B$ also living in the braided
category $\cal C$. It has product
\begin{equation}
\matrix{\object{A\!\underline\otimes\!B}
               \Step\object{A\!\!\underline\otimes\!\!B}\cr
        \vvbox{\hbox{\cu}}\cr
        \step\object{A\!\underline\otimes\!B}\step}
\quad ={}\quad{}
(\cdot_A\otimes\cdot_B)\circ (\hbox{id}\otimes\Psi_{B,A}\otimes\hbox{id})
\quad =\quad
\matrix{\object{A}\Step\object{B}\Step\object{A}\Step\object{B}\cr
	\vvbox{\hbox{\id\Step\x\Step\id}
               \hbox{\cu\Step\cu}}\cr
        \step\object{A}\Step\Step\object{B}\step}
\end{equation}
\end{proposition}

\begin{definition}
\begin{itemize}
\item
{\em Bialgebra $B$ in braded category $\cal C$} is
object in $\cal C$ equipped with algebra and coalgebra structures
obeying the following compatibility axioms
\begin{equation}
\vvbox{\hbox{\cu}
       \hbox{\cd}}
\quad ={}\quad{}
\vvbox{\hbox{\cd\step\cd}
       \hbox{\id\Step\hx\Step\id}
       \hbox{\cu\step\cu}}
\label{bialgebra}
\qquad\qquad
\vvbox{\hbox{\step\unit}\hbox{\cd}}
\quad ={}\quad{}
\vvbox{\hbox{\unit\step\unit}\hbox{\id\step\id}}
\qquad\qquad
\vvbox{\hbox{\cu}\hbox{\step\counit}}
\quad ={}\quad{}
\vvbox{\hbox{\id\step\id}\hbox{\counit\step\counit}}
\qquad\qquad
\vvbox{\hbox{\unit}
       \hbox{\counit}}
\quad ={}\quad{}
\end{equation}
(i.e. $\Delta :\, B\rightarrow B\underline\otimes B$,
      $\epsilon :\, B\rightarrow\underline 1$
are algebra homomorphisms or
     $B\underline\otimes B\rightarrow B$,
     $\eta :\, \underline 1\rightarrow B$
are coalgebra homomorphisms.)

\item
{\em Hopf algebra $H$ in braded category $\cal C$ (braded Hopf algebra)} is
bialgebra in $\cal C$ with antipode $S:\,H\rightarrow H$ which is
convolution-inverse to identical map:
\begin{equation}
\vvbox{\hbox{\cd}\hbox{\S\Step\id}\hbox{\cu}}
\quad ={}\quad{}
\vvbox{\hbox{\counit}\hbox{\unit}}
\quad ={}\quad{}
\vvbox{\hbox{\cd}\hbox{\id\Step\S}\hbox{\cu}}
\end{equation}
We suppose also that there exists inverse morphism $S^{-1}$.

It follows from Hopf algebra axioms that antipode is anti-algebra and
anti-coalgebra map \cite{Majid8}:
\begin{equation}
\vvbox{\hbox{\cu}
       \hbox{\step\S}}
\quad ={}\quad{}
\vvbox{\hbox{\S\Step\S}
       \hbox{\x}
       \hbox{\cu}}
\qquad\qquad
\vvbox{\hbox{\step\S}
       \hbox{\cd}}
\quad ={}\quad{}
\vvbox{\hbox{\cd}
       \hbox{\x}
       \hbox{\S\Step\S}}
\end{equation}
\end{itemize}
\end{definition}

\begin{remark}
Definition of usual Hopf algebra \cite{Sweedler1} is directly extended on
the braided case.

We say nothing about additive structure. For example, category of sets with
Cartesian product is tensor one and Hopf algebras in this category are
exactly usual groups.

All results of this paper hold if we suppose all categories and functors
to be $k$-linear, where $k$ is commutative ring.
\end{remark}

\begin{proposition} \cite{Majid8}:
If $H$ is a braided-Hopf algebra with left dual object ${}^\vee H$ then
${}^\vee H$ is also braided-Hopf algebra with product, coproduct, antipode,
unit and counit given by
\begin{equation}
\matrix{\object{{}^\vee H}\Step\object{{}^\vee H}\cr
	\vvbox{\hbox{\cu}}\cr
        \step\object{{}^\vee H}\step}
\quad ={}\quad{}
{\unitlength =1.2\unitlength
\vcenter{\hbox{\begin{picture}(2.5,4)
	 \put(0,2){\line(0,1){2}}
	 \put(.5,2){\line(0,1){2}}
	 \put(1.75,2.5){\line(0,1){.5}}
	 \put(2.5,0){\line(0,1){3}}
	 \put(1,2){\oval(1,1)[b]}
	 \put(1,2){\oval(2,2)[b]}
	 \put(1.75,2){\oval(.5,.5)[t]}
	 \put(2.125,3){\oval(.75,.75)[t]}
         \end{picture}}}}
\qquad\qquad
\matrix{\step\object{{}^\vee H}\step\cr
	\vvbox{\hbox{\cd}}\cr
        \object{{}^\vee H}\Step\object{{}^\vee H}}
\quad ={}\quad{}
{\unitlength =1.2\unitlength
\vcenter{\hbox{\begin{picture}(2.5,4)
	 \put(2,0){\line(0,1){2}}
	 \put(2.5,0){\line(0,1){2}}
	 \put(.75,1){\line(0,1){.5}}
	 \put(0,1){\line(0,1){3}}
	 \put(1.5,2){\oval(1,1)[t]}
	 \put(1.5,2){\oval(2,2)[t]}
	 \put(.75,2){\oval(.5,.5)[b]}
	 \put(.375,1){\oval(.75,.75)[b]}
         \end{picture}}}}
\end{equation}
\begin{equation}
\matrix{\object{{}^\vee H}\cr
	\vvbox{\hbox{\S}}\cr
        \object{{}^\vee H}}
\quad ={}\quad{}
\vvbox{\hbox{\id\step\hcoev}
       \hbox{\id\step\S\step\id}
       \hbox{\hev\step\id}}
\qquad\qquad
\matrix{\vvbox{\hbox{\unit}}\cr
	\object{{}^\vee H}}
\quad ={}\quad{}
\vvbox{\hbox{\hcoev}
       \hbox{\counit\step\id}}
\qquad\qquad
\matrix{\object{{}^\vee H}\cr
        \vvbox{\hbox{\counit}}}
\quad ={}\quad{}
\vvbox{\hbox{\id\step\unit}
       \hbox{\hev}}
\end{equation}
\end{proposition}

\subsection{Symmetries.}

Let $\cal C$ is braided monoidal category.
Then one can build another ones:
\begin{itemize}
\item
$\cal C _{\rm sym}$ the the same category with symmetrical tensor product
\begin{displaymath}
X\otimes _{\rm sym} Y:=Y\otimes X
\end{displaymath}
and braiding
\begin{displaymath}
X\otimes _{\rm sym} Y=Y\otimes X  \buildrel{\Psi _{Y,X}}\over\longrightarrow
  X\otimes Y=Y\otimes _{\rm sym} X.
\end{displaymath}
\item
${\cal C}^{\rm op}$ is opposite category with braiding
\begin{displaymath}
X\otimes Y \buildrel {\Psi _{Y,X}} \over \longleftarrow Y\otimes X.
\end{displaymath}
\item
$\overline{\cal C}$ is the same category with the same tensor product and with
mirror-reversed braiding
\begin{displaymath}
\overline{\Psi_{X,Y}}={\Psi_{Y,X}}^{-1}
\end{displaymath}
\end{itemize}

\begin{remark}
Identical functor with natural transformation
$\Psi_{X,Y}:\;X\otimes Y\rightarrow X\otimes_{\rm sym}Y$
is isomorphism of braided categories $\cal C$ and ${\cal C}_{\rm sym}$.
\label{sym}
\end{remark}

For diagrammatic form of map in $\cal C$ choose coordinate system with
$X$-axis (resp. $Z$-axis) horizontal (resp. vertical) in the plane of the
drawing and $Y$-axis orthogonal to this plane.

Then diagrammatic forms of map in $\cal C$ and corresponding map in
$\overline{\cal C}$ (resp. ${\cal C}^{\rm op}$, ${\cal C}^{\rm op}$)
are obtained one from other by
the plane $XZ$-symmetry ({\em mirror symmetry})
(resp.
axial $X$-symmetry ({\em input-output symmetry}),
axial $Z$-symmetry ({\em left-rigt symmetry})).
One can apply these symmetries to any proof which has a form of identities
between diagrams.
So symmetries of axiom imply symmetries of theorems.

Notice that collection of Hopf algebra axioms for $H$ in braided category
$\cal C$ is input-output and left-rigt symmetrical,
i.e. $H$ is Hopf algebra in categories ${\cal C}_{\rm sym}$ and
${\cal C}_{\rm op}$ also.

\interskip
For Hopf agebra $H$ in $\cal C$ denote by $H^{\rm op}$ (resp. $H_{\rm op}$)
the same coalgebra (resp. algebra) with opposite multiplication
$\cdot^{\rm op}$ (resp. comultiplication $\Delta^{\rm op}$):
\begin{equation}
\cdot^{\rm op}\quad :=\quad
\vvbox{\hbox{\hxx}
       \hhbox{\cu}}
\qquad\qquad
\Delta^{\rm op}\quad :=\quad
\vvbox{\hhbox{\cd}
       \hbox{\hxx}}
\end{equation}
\begin{proposition}(\cite{Majid8})
$H^{\rm op}$ and $H_{\rm op}$ are Hopf algebras in $\overline{\cal C}$ with
antipode $S^{-1}$.
\label{opHopf}
\end{proposition}

\begin{remark}
\begin{itemize}
\item
If $H$ has dual ${}^\vee H$ then
${}^\vee (H^{\rm op})$ and $({}^\vee H)_{\rm op}$ are the same Hopf algebras.
\item
$(H^{\rm op})_{\rm op}
\buildrel S\over\longrightarrow H \buildrel S\over\longrightarrow
(H_{\rm op})^{\rm op}$
are isomorphisms of Hopf algebras in category $\cal C$.
\item
We have the following pair of mutually inverse isomorphisms of Hopf algebras
\begin{displaymath}
\matrix{\object{(({}^\vee H)^{\rm op})_{\rm op}}\Step\step\cr
	\vvbox{\hbox{\id\step\coev}
	       \hbox{\hx\Step\id}
	       \hbox{\id\step\ev}}\cr
        \object{((H^\vee )_{\rm op})^{\rm op}}\Step\step}
\qquad{\rm and}\qquad
\matrix{\Step\step\object{((H^\vee)_{\rm op})^{\rm op}}\cr
	\vvbox{\hbox{\coev\step\id}
	       \hbox{\id\Step\hxx}
	       \hbox{\ev\step\id}}\cr
        \Step\step\object{(({}^\vee H)^{\rm op})_{\rm op}}}
\end{displaymath}
\end{itemize}
\label{oppdual}
\end{remark}

\subsection{Modules. Comodules.}

\begin{definition}
Object $X$ is {\em right module over algebra} $A$ in monoidal category $\cal C$
if action  $\triangleleft :\;X\otimes A\rightarrow X$
satisfys axioms akin to (\ref{algebra})
(We use the following notation for action to differ it from multiplication.):
\begin{equation}
\triangleleft
\quad :={}\quad
\matrix{\object{X}\step\object{A}\cr
         \vvbox{\hbox{\ru}}\cr
         \object{X}\step \cr}
\qquad\qquad
\vvbox{\hbox{\id\step\unit}
       \hbox{\ru}}
\quad ={}\quad
\vvbox{\hbox{\id}
       \hbox{\id}}
\qquad\qquad
\vvbox{\hbox{\ru\step\id}
       \hbox{\Ru}}
\quad ={}\quad
\vvbox{\hbox{\id\step\cu}
       \hbox{\Ru}}
\end{equation}
\end{definition}

\begin{proposition}
\begin{itemize}
\item
Category ${\cal C}_H$ of right modules over Hopf algebra $H$ in braided
category $\cal C$ is monoidal with the following module structure on
tensor product $X\otimes Y$ of modules $X$ and $Y$:
\begin{equation}
\matrix{\object{X}\step\object{Y}\hstep\step\object{H}\hstep\cr
	\vvbox{\hhbox{\krrr\id\step\id\step\cd}
	       \hbox{\id\step\hx\step\id}
	       \hbox{\ru\step\ru}}\cr
	\object{X}\Step\object{Y}\step}
\label{tensorproduct}
\end{equation}
and with associativity isomorphisms inherited from initial category $\cal C$.
\item
Right (resp. left) dual for module $X$ in ${\cal C}_H$ is right (resp. left)
dual for its underlying object in $\cal C$ (if this latter exists) with the
action
\begin{equation}
\matrix{\Step\object{X^\vee}\step\object{H}\cr
        \vvbox{\hbox{\Step\id\step\S}
               \hbox{\hcoev\step\hx}
               \hbox{\id\step\ru\step\id}
               \hbox{\id\step\ev}}\cr
	\object{X^\vee}\Step\step}
\qquad\qquad
\matrix{\object{{}^\vee X}\step\object{H}\Step\cr
        \vvbox{\hbox{\id\step\SS\step\hcoev}
               \hbox{\id\step\hxx\step\id}
               \hbox{\id\ru\step\id\step\id}
               \hbox{\ev\step\id}}\cr
	\Step\step\object{{}^\vee X}}
\label{dualmodule}
\end{equation}
\end{itemize}
\end{proposition}

\begin{remark}
Analogous result for comodules is obtained using input-output symmetry.
\end{remark}

\section{Crossed modules.}

Notion of crossed module over usual Hopf algebra belongs to Yetter
\cite{Yetter1}.
In this section we describe braded categories
${\cal C}^H_H$ and ${}^H_H{\cal C}$
of crossed modules over Hopf algebra $H$ in braded category $\cal C$.

\begin{definition}
{\em Right crossed module} $X$ over Hopf algebra $H$ in braded category
$\cal C$ is object with right module and comodule structures over $H$ obeying
compatibility axiom
\begin{equation}
\matrix{\object{X}\hstep\step\object{H}\step\cr
        \vvbox{\hhbox{\krrr\id\step\cd}
               \hbox{\hx\step\id}
               \hhbox{\krrr\id\step\ru}
               \hbox{\id\step\rd}
               \hbox{\hx\step\id}
               \hhbox{\krrr\id\step\cu}}\cr
       \object{X}\hstep\step\object{H}\step}
\quad ={}\quad
\matrix{\object{X}\Step\step\object{H}\step\cr
        \vvbox{\hbox{\rd\step\cd}
	       \hhbox{\krrr\id\step\id\step\id\Step\id}
               \hbox{\id\step\hx\Step\id}
	       \hhbox{\krrr\id\step\id\step\id\Step\id}
	       \hbox{\ru\step\cu}}\cr
       \object{X}\Step\step\object{H}\step}
\label{crossedmodule}
\end{equation}

${\cal C}_H^H$ is the {\em category of right crossed modules} with morphisms
which are both module and comodule maps.
\end{definition}

\begin{example}
Underlying object of a Hopf algebra $H$ in braided category $\cal C$
with adjoint action
${}_{\rm ad}\triangleleft$
and regular coaction
${}^{\rm reg}\beta :=\Delta$
\begin{equation}
{}_{\rm ad}\triangleleft\quad :=\quad
\matrix{\hstep\object{H_{\rm ad}}\step\hstep\object{H}\hstep\cr
        \vvbox{\hhbox{\krrr\hstep\id\step\cd}
               \hbox{\hstep\hx\step\id}
	       \hhbox{\krrr\dd\step\id\step\id}
               \hbox{\S\step\hstep\hcu}
	       \hbox{\cu}}\cr
	\step\object{H_{\rm ad}}\step\hstep}
\qquad\qquad
{}^{\rm reg}\beta\quad :=\quad
\matrix{\step\object{H_{\rm ad}}\step\cr
        \vvbox{\hbox{\cd}}\cr
	\object{H_{\rm ad}}\Step\object{H}}
\end{equation}
is right crossed module over $H$.

Using input-output symmetry one can construct another crossed module
$H^{\rm ad}$ with adjoint coaction ${}^{\rm ad}\beta$ and regular action
${}_{\rm reg}\triangleleft :=\cdot$.
\end{example}

\begin{lemma}
If $X$ and $Y$ are crossed modules over braided Hopf algebra $H$ then
$X\otimes Y$ is the same with module and comodule structure defined by
the diagram (\ref{tensorproduct}) and input-output symmetrical one.
\end{lemma}

\begin{proof}
See figures \ref{prooftensorproduct1}-\ref{prooftensorproduct2}.
\end{proof}

Notice that if $f_i:\,X_i\rightarrow Y_i,\enspace i=1,2$ are crossed module
maps then
$f_1\otimes f_2:\,X_1\otimes X_2\rightarrow Y_1\otimes Y_2$
is the same.
So ${\cal C}^H_H$ is monoidal category.
Unit object is one in $\cal C$ with trivial crossed module structure.

\begin{theorem}
Category ${\cal C}^H_H$ is braided with
\begin{equation}
{}^{({\cal C}^H_H)}\Psi
\quad ={}\quad
\vvbox{\hbox{\id\step\rd}
       \hbox{\hx\step\id}
       \hbox{\id\step\ru}}
\qquad\qquad\qquad
{}^{({\cal C}^H_H)}\Psi^{-1}
\quad ={}\quad
\vvbox{\hbox{\rd\step\id}
       \hbox{\id\step\hxx}
       \hbox{\hxx\step\SS}
       \hbox{\id\step\hxx}
       \hbox{\ru\step\id}}
\label{crossedmodulePsi}
\end{equation}
\end{theorem}

\begin{proof} See figures \ref{proofmodulemap}--\ref{proofinverse}.\end{proof}

\begin{proposition}
Right dual for crossed module $X$ in ${\cal C}_H^H$ is right dual for its
underlying object in $\cal C$ (if this latter exists) with the action and
coaction
\begin{equation}
\matrix{\Step\object{X^\vee}\step\object{H}\cr
        \vvbox{\hbox{\hcoev\step\id\step\S}
               \hbox{\id\step\id\step\hx}
               \hbox{\id\step\ru\step\id}
               \hbox{\id\step\ev}}\cr
       \object{X^\vee}\Step\step}
\qquad\qquad
\matrix{\Step\step\object{X^\vee}\cr
        \vvbox{\hbox{\hcoev\Step\id}
               \hbox{\id\step\rd\step\id}
               \hbox{\id\step\hx\step\id}
               \hbox{\id\step\SS\step\hev}}\cr
	\object{X^\vee}\step\object{H}\Step}
\end{equation}
\end{proposition}

\begin{proof}
Module (resp. comodule) structure on $X^\vee$ is like in category
${\cal C}_H$ (\ref{dualmodule})(resp. ${\cal C}^H$).
Evaluation and coevaluation are module and comodule maps like in categories
${\cal C}_H$ and ${\cal C}^H$ respectively.
Verification of crossed module axiom see on figures
\ref{proofdual1}-\ref{proofdual2}.
\end{proof}

Analog of this proposition for left dual ${}^\vee X$ are obtained
using input-output symmetry.

\interskip
The definitions and results like previous are true for category
${}^H_H{\cal C}$ of left crossed modules. One needs to substitute all
diagramms for left-right symmetrical.

\interskip
Monoidal categories ${\cal C}_H$ and ${}_H{\cal C}$ of right and left
modules over Hopf algebra $H$ in $\cal C$ are isomorphic.
We will use a similar construction to prove the following:

\begin{theorem}
For Hopf algebra $H$ in braided category $\cal C$ braided categories
${\cal C}_H^H$ and ${}_H^H{\cal C}$ are isomorphic.
\label{iso}
\end{theorem}

There exist two canonical constructions of left crossed module
from right crossed module $X$.
One can use any of them to build isomorphism mentioned above.
Denote by $X^S$ (resp. ${}^SX$) underlying object of $X$ with the left action
and coaction defined by the formulas
\begin{equation}
X^S\;:\qquad
\matrix{\object{H}\step\object{X^S}\cr
        \vvbox{\hbox{\SS\step\id}
               \hbox{\hxx}
               \hbox{\ru}}}
\qquad
\matrix{
        \vvbox{\hbox{\rd}
               \hbox{\hx}
               \hbox{\S\step\id}}\cr
	\object{H}\step\object{X^S}}
\qquad\qquad
{}^SX\;:\qquad
\matrix{\object{H}\step\object{{}^SX}\cr
        \vvbox{\hbox{\S\step\id}
               \hbox{\hx}
               \hbox{\ru}}}
\qquad
\matrix{
        \vvbox{\hbox{\rd}
               \hbox{\hxx}
               \hbox{\SS\step\id}}\cr
	\object{H}\step\object{{}^SX}}
\end{equation}

\begin{lemma}
$X^S$ and ${}^SX$ are left crossed modules over $H$.
\label{lemma1}
\end{lemma}

\begin{proof} See figure \ref{proofleftright}. \end{proof}

Next lemma show that functor
${\rm Obj}({\cal C}^H_H)\ni X\rightarrow X^S\in{\rm Obj}({}^H_H{\cal C})$
identical on morphisms together with natural transformation
\begin{equation}
(X\otimes Y)^S
\buildrel{{}^{\cal C}\Psi}\over\longrightarrow
X^S\otimes_{\rm sym}Y^S=Y^S\otimes X^S
\label{nattrans}
\end{equation}
is isomorphism of braided categories ${\cal C}$ and $({\cal C})_{\rm sym}$.
Taking into attention remark \ref{sym} this prove theorem \ref{iso}.

\begin{lemma}
Let $X$ and $Y$ are right crossed modules over $H$. Then
\begin{itemize}
\item
(\ref{nattrans}) is crossed module isomorphism;
\item
The following diagram is commutative
\begin{displaymath}
\matrix{(X\otimes Y)^S=X\otimes Y&
   \!\!\buildrel{{}^{\overline{({\cal C}^H_H)}}\Psi}\over\longrightarrow\!\!&
	(Y\otimes X)^S=Y\otimes X\cr\cr
	\downarrow^{{}^{\overline{\cal C}}\Psi}&&
	\downarrow^{{}^{\overline{\cal C}}\Psi}\cr
        X^S\otimes_{\rm sym}Y^S=Y\otimes X&
  \!\!\buildrel{{}^{({}^H_H{\cal C})}\Psi}\over\longrightarrow\!\!&
	Y^S\otimes_{\rm sym}X^S=X\otimes Y}
\end{displaymath}
\end{itemize}
\label{lemma2}
\end{lemma}

\begin{definition}
For right crossed module $X$ define the following morphisms in
$\cal C$ (but not in ${\cal C}^H_H$):
\begin{equation}
\matrix{\object{X}\cr
	\vvbox{\hbox{\O{S^2}}}\cr
	\object{X}}
\quad :=\quad
\vvbox{\hbox{\rd}
       \hbox{\id\step\S}
       \hbox{\ru}}
\qquad\qquad
\matrix{\object{X}\cr
	\vvbox{\hbox{\O{S^{-2}}}}\cr
	\object{X}}
\quad :=\quad
\vvbox{\hbox{\rd}
       \hbox{\hxx}
       \hbox{\O{S^{-2}}\step\id}
       \hbox{\hxx}
       \hbox{\ru}}
\label{squareantipode}
\end{equation}
(We consider $S^2$ as one symbol but not as a square of $S$
which is not defined on crossed modules.)
\end{definition}

The following proposition justify our notations

\begin{proposition}
For crossed modules $H_{\rm ad}$ and $H^{\rm ad}$ morphisms defined by
(\ref{squareantipode}) coincide with square of usual antipode and its inverse.
\end{proposition}

\begin{proposition}
Map $X^S\buildrel{S^2}\over\longrightarrow{}^SX$
is isomorphism between left crossed modules $X^S$ and ${}^SX$ with inverse
${}^SX\buildrel{S^{-2}}\over\longrightarrow X^S$.
\end{proposition}

\begin{proof}
See figures \ref{proofantipodemodulemap}-\ref{proofantipodeinvert}.
\end{proof}

\begin{proposition}
The following identity is true for any crossed modules $X$ and $Y$ from
${\cal C}_H^H$:
\begin{equation}
{}^{({\cal C}^H_H)}\Psi_{Y,X}\circ S^2\circ{}^{({\cal C}^H_H)}\Psi_{X,Y}=
{}^{\cal C}\Psi_{Y,X}\circ (S^2\otimes S^2)\circ
                                                 {}^{\cal C}\Psi_{X,Y}
\label{antipodetensor}
\end{equation}
\end{proposition}

\begin{proof}
See figure \ref{proofantipodetensor}.
\end{proof}

We shall return to the last formula when consider quantum braded groups.

\begin{definition}
For object $X$ in braided category $\cal C$ which has left and right dual
${}^\vee X$, $X^\vee$ rang \cite{Majid7} or $8$-dimension \cite{Lyubashenko5}
is element of ${\rm End}(\underline 1)$
\begin{equation}
{\rm dim}^{\cal C}_8(X)
\quad :=\quad
\vvbox{\hbox{\obj{X}\coev}
       \hbox{\x}
       \hbox{\ev}}
\end{equation}
\end{definition}

\begin{proposition}
Let $X$ be crossed module with duals ${}^\vee X$, $X^\vee$. Then
\begin{equation}
\matrix{\object{X}\Step\object{X^\vee}\cr
	  \vvbox{\hbox{\O{S^2}\Step\id}
		 \hbox{\ev}}}
\quad =\quad
\matrix{\object{X}\Step\object{X^\vee}\cr
	  \vvbox{\hbox{\id\Step\O{S^2}}
		 \hbox{\ev}}}
\qquad\qquad
{\rm dim}^{{\cal C}^H_H}_8(X)
\quad :=\quad
\vvbox{\hbox{\coev}
       \hbox{\O{S^2}\Step\id}
       \hbox{\x}
       \hbox{\ev}}
\quad =\quad
\vvbox{\hbox{\coev}
         \hbox{\id\Step\O{S^2}}
	 \hbox{\x}
	 \hbox{\ev}}
\end{equation}
\end{proposition}

\interskip
There exists natural connection between cetegories of crossed modules over
Hopf algebra and over its opposite or dual.

\begin{theorem}
Let $H$ be a Hopf algebra in braided category $\cal C$.
Then there exist the following isomorphisms between braided categories:
\begin{enumerate}\begin{enumerate}
\item
$({\cal C}^{\rm op})^H_H\simeq ({\cal C}_H^H)^{\rm op}$
\item
$\overline{({\cal C}_H^H)}\simeq
 \overline{\cal C}^{H_{\rm op}}_{H_{\rm op}}\simeq
 \overline{\cal C}^{H^{\rm op}}_{H^{\rm op}}$
\item
${\cal C}^H_H
 \simeq{\cal C}^{H^\vee}_{H^\vee}
 \simeq{\cal C}^{\,{}^\vee\! H}_{\,{}^\vee\! H}\quad$
(if $H^\vee$ or ${}^\vee H$ exist).
\end{enumerate}\end{enumerate}
\end{theorem}

({\it a}) is obvious: all diagrams in definitions are input-output symmetrical.
Proof of ({\it b}) and ({\it c}) is like in theorem \ref{iso} :
we prove equivalent facts
$\overline{({\cal C}_H^H)}\simeq
 (\overline{\cal C}^{H_{\rm op}}_{H_{\rm op}})_{\rm sym}\,,\enspace
 {\cal C}^H_H\simeq
 ({\cal C}^{\,{}^\vee\!H}_{\,{}^\vee\!H})_{\rm sym}$
(for $H^{\rm op}$ and $H^\vee$ one needs to consider input-output symmetrical
diagrams or use remark \ref{oppdual}.

Let $X$ be right crossed module over $H$.
Then $X_{[{\rm op}]}$ (resp. ${}^{[\vee ]}X$)
is right module and comodule over $H_{\rm op}$ (resp. ${}^\vee H$)
with the same underlying object and the following action and coaction
\begin{equation}
\vvbox{\hbox{\ru}}\qquad\quad
\vvbox{\hbox{\rd}
       \hbox{\hxx}
       \hbox{\SS\step\id}
       \hbox{\hxx}}
\qquad\hbox{for}\enspace X_{[{\rm op}]}
\qquad\qquad\hbox{and}\qquad\qquad
\vvbox{\hbox{\hx}
       \hbox{\id\step\rd}
       \hbox{\hx\step\id}
       \hbox{\id\step\hev}}
\qquad\quad
\vvbox{\hbox{\id\step\hcoev}
       \hbox{\ru\step\id}}
\qquad\hbox{for}\enspace {}^{[\vee ]}X
\end{equation}

\begin{lemma}
$X_{[{\rm op}]}$ and ${}_{[{\rm op}]}X$ are crossed modules.
\end{lemma}

\begin{proof} See figures \ref{proofcop1} -- \ref{proofvee2}. \end{proof}

\begin{lemma}
Let $X$ and $Y$ are right crossed modules over $H$ then
\begin{itemize}
\item
There exist the following isomorphisms of crossed modules
\begin{displaymath}
(X\otimes Y)_{[{\rm op}]}
\buildrel{{}^{\overline{\cal C}}\Psi}\over\longrightarrow
X_{[{\rm op}]}\otimes_{\rm sym}Y_{[{\rm op}]}
\end{displaymath}
\begin{displaymath}
{}^{[\vee ]}(X\otimes Y)
\buildrel{{}^{\cal C}\Psi}\over\longrightarrow
{}^{[\vee ]}X\otimes_{\rm sym}{}^{[\vee ]}Y
\end{displaymath}
\item
The following diagrams are commutative
\begin{displaymath}
\matrix{X\otimes Y&
   \!\!\buildrel{{}^{\overline{({\cal C}^H_H)}}\Psi}\over\longrightarrow\!\!&
	Y\otimes X\cr\cr
	\downarrow^{{}^{\overline{\cal C}}\Psi}&&
	\downarrow^{{}^{\overline{\cal C}}\Psi}\cr
        Y\otimes X&
  \!\!\buildrel{{}^{(\overline{\cal C}^{H_{\rm op}}_{H_{\rm op}})}\Psi}
                 \over\longrightarrow\!\!&
	X\otimes Y}
\qquad\qquad
\matrix{X\otimes Y&
	\!\!\buildrel{{}^{{\cal C}^H_H}\Psi}\over\longrightarrow\!\!&
	Y\otimes X\cr\cr
	\downarrow^{{}^{\cal C}\Psi}&&
	\downarrow^{{}^{\cal C}\Psi}\cr
        Y\otimes X&
 \!\!\buildrel{{}^{({\cal C}^{H^\vee}_{H^\vee})}\Psi}\over\longrightarrow\!\!&
	X\otimes Y}
\end{displaymath}
\end{itemize}
\end{lemma}

\interskip
Braided category ${}_H^H{\cal M}$ of crossed modules over usual Hopf algebra
$H$ can be obtained directly from monoidal categry ${}_H{\cal M}$ as
a 'center' or 'inner double'.
In the rest of this section we show that the same with slight modification
is true in braided case.
The following construction is a special case of {\em Pontryagin dual monoidal
category} \cite{Majid9} (existence of braiding in this case was pointed out
by Drinfeld).

\begin{proposition} \cite{Majid9} \cite{Majid10}
Let ${\cal C}$ be a monoidal category. There is a braided monoidal category
$\cal Z(C)$ ('center' or 'inner double' of $\cal C$) defined as follows.
Objects are pairs $(V,\lambda_V)$ where $V$ is an object of $\cal C$ and
$\lambda_V$ is a natural isomorfism in
$\hbox{Nat}(V\otimes\hbox{id},\hbox{id}\otimes V)$ such that
$$\lambda_{V,\underline 1}={\rm id}\qquad
  (\hbox{id}\otimes\lambda_{V,Z})(\lambda_{V,W}\otimes\hbox{id})=
  \lambda_{V,W\otimes Z}$$
and morphisms are $\phi :\; V\rightarrow W$ such that the objects are
intertwined in the form
$$(\hbox{id}\otimes\phi )\lambda_{V,Z}=
  \lambda_{W,Z}(\phi\otimes\hbox{id}\,,\qquad
  \forall Z \in{\rm Obj}({\cal C})$$
The tensor product and braiding are
$$(V,\lambda_V)\otimes (W,\lambda_W)=
  (V\otimes W,\lambda_{V\otimes W})\qquad
  \lambda_{V\otimes W,Z}=
  (\lambda_{V,Z}\otimes\hbox{id})(\hbox{id}\otimes\lambda_{W,Z})$$
$$\Psi_{(V,\lambda_V),(W,\lambda_W)}=\lambda_{V,W}$$
\end{proposition}

\begin{proposition} \cite{Majid10}
Let $H$ be usual Hopf algebra over field $k$ and ${}_H{\cal M}$ the monoidal
category of $H$-modules.
Then ${\cal Z}({}_H{\cal M})$ coincides with the category ${}_H^H{\cal M}$
of Yetter's crossed modules \cite{Yetter1}
\label{prop}
\end{proposition}

For Hopf algebra $H$ in braided category $\cal C$ denote by
${\cal Z}_{\cal C}({}_H{\cal C})$ full subcategory of ${\cal Z}({}_H{\cal C})$
with condition on $\lambda_V$ as for any object $W$ in $\cal C$ with trivial
action (throught counit) $\lambda_{V,W}$ coincides with braiding
${}^{\cal C}\Psi_{V,W}$ in $\cal C$.
(In tensor category of vector spaces Vect unit object $\underline 1=k$ is
generator then ${\cal Z}_{\rm Vect}({}_H{\cal M})={\cal Z}({}_H{\cal M})$.)

\begin{proposition}
Braided monoidal categories
${}_H^H{\cal C}$ and ${\cal Z}_{\cal C}({}_H{\cal C})$ are isomorphic.
\end{proposition}

\begin{proof}
Proof is like in proposition \ref{prop}.
One can identify crossed module $X$ with pair: underlying module of $X$ and
braiding ${}^{({}^H_H{\cal C})}\Psi_{X,\_}$.
Conversely, coaction on $X$ can be reconstructed from $(X,\lambda_X)$ as
composition
$X\buildrel{{\rm id}\otimes\eta}\over\longrightarrow X\otimes H
  \buildrel{\lambda_X}\over\longrightarrow H\otimes X$.
(We conside $H$ with left regular module structure.)
\end{proof}

Similarly, one can identify ${\cal C}^H_H$ and ${\cal Z}_{\cal C}({\cal C}^H)$.
So 'centers' of categories of modules and comodules are the same and coinside
with category of crossed modules.

\section{Cross products.}

\begin{proposition}
\label{adjointalgebra}
Let $A$ be a Hopf algebra in braided category $\cal C$.
Then $A_{\rm ad}$ (resp. $A^{\rm ad}$) become commutative algebra
(resp. cocommutative coalgebra) in category ${\cal C}^A_A$ with
multiplication (resp. comultiplication) inherited from $A$.
\end{proposition}

Let $B$ be a Hopf algebra in category ${\cal C}_A^A$.
Denote by $A\ltimes B$ object $A\otimes B$ equipped with
algebra structure $A_{\rm ad}\underline\otimes B$,
coalgebra structure $A^{\rm ad}\underline\otimes B$
(tensor product algebra and coalgebra in ${\cal C}_A^A$) and antipode
$$S_{A\ltimes B}:=
  {}^{({\cal C}_A^A)}\Psi_{B,A_{\rm ad}}\circ (S_B\otimes S_A)\circ
                             {}^{({\cal C}_A^A)}\Psi_{A^{\rm ad},B}.$$
(It's easy to see that there exists ${S_{A\ltimes B}}^{-1}$ because all
factors in the last formula are invertible.)
In terms of category $\cal C$ multiplication, comaltiplication and antipode
are the following:
\begin{equation}
\matrix{\object{A}\step\object{B}\step\hstep\object{A}\step\hstep\object{B}\cr
	\vvbox{\hbox{\id\step\id\step\hcd\step\id}
	       \hbox{\id\step\hx\step\id\step\id}
	       \hbox{\id\step\id\step\ru\step\id}
	       \hbox{\hcu\step\cu}}\cr
	\hstep\object{A}\hstep\Step\object{B}\step}
\qquad\qquad
\matrix{\hstep\object{A}\hstep\Step\object{B}\step\cr
	\vvbox{\hbox{\hcd\step\cd}
	       \hbox{\id\step\id\step\rd\step\id}
	       \hbox{\id\step\hx\step\id\step\id}
               \hbox{\id\step\id\step\hcu\step\id}}\cr
        \object{A}\step\object{B}\step\hstep\object{A}\step\hstep\object{B}}
\qquad\qquad
\matrix{\object{A}\step\object{B}\step\cr
	\vvbox{\hbox{\id\step\rd}
	       \hbox{\hx\step\id}
	       \hhbox{\id\step\cu}
	       \hbox{\S\step\hstep\S}
	       \hhbox{\id\step\cd}
	       \hbox{\hx\step\id}
	       \hbox{\id\step\ru}}\cr
        \object{A}\step\object{B}\step}
\label{CrossProd}
\end{equation}

\begin{theorem}
$A\ltimes B$ is a Hopf algebra in category $\cal C$.
\label{CrossProduct}
\end{theorem}

\begin{proof}
Nontrivial part is verification of bialgebra axiom
(see figures \ref{proofbialgebra1}-\ref{proofbialgebra3}).
\end{proof}

Analog of Radford's theorem (\cite{Radford1}; \cite{Majid10}, proposition 4.15)
is also true in this general situation.
Previously we formulate some additional assumption about category $\cal C$
and show that it can be always satisfyed.

\begin{definition}
We say that idempotent (projection) $e=e^2:\,X\rightarrow X$ is split in
category $\cal C$ if there exist object $X_e$ and maps
$X_e \matrix{i_e\cr \longrightarrow\cr \longleftarrow\cr p_e} X$
such that $e=i_e\circ p_e$ and $p_e\circ i_e={\rm id}_{X_e}$.
\end{definition}

\begin{proposition}
For any braided category $\cal C$ there exists full imbedding of braided
categories ${\cal C}\hookrightarrow\widetilde{\cal C}$
and idempotents in $\widetilde{\cal C}$ are split.
\end{proposition}

\begin{proof}
We need only to prove that standard construction of idempotent spliting
is compatible with braided structure. Objects in category
$\widetilde{\cal C}$ are pairs $X_e=(X,e)$, where $X$ is object in $\cal C$
and $e:\,X\rightarrow X$ is idempotent: $e^2=e$. Morphisms are the followng
$$\widetilde{\cal C}(X_e,Y_f):=\{t\in{\cal C}(X,Y)\bracevert fte=t\}$$
with usual composition.
Tensor product, associativity morphisms and brading are:
$$X_e\otimes Y_f:=(X\otimes Y)_{e\otimes f}$$
$$\Phi_{X_e,Y_f,Z_g}:=
  ((e\otimes f)\otimes g)\circ\Phi_{X,Y,Z}\circ (e\otimes (f\otimes g))=
  ((e\otimes f)\otimes g)\circ\Phi_{X,Y,Z}=
  \Phi_{X,Y,Z}\circ (e\otimes (f\otimes g))$$
$$\Psi_{X_e,Y_f}:=
  (f\otimes e)\circ\Psi_{X,Y}\circ (e\otimes f)=
  (f\otimes e)\circ\Psi_{X,Y}=
  \Psi_{X,Y}\circ (e\otimes f)$$
Axioms of braided category are easily verified.
One can identify $\cal C$ with full subcategory of $\widetilde{\cal C}$
whose objects are $(X,{\rm id}_X)$.
\end{proof}

\begin{proposition}
Let $\cal C$ be a braided category with split idempotents and $H$ a Hopf
algebra in $\cal C$. Then idempotents in categories ${\cal C}_H$, ${\cal C}^H$
and ${\cal C}_H^H$ are split.
\end{proposition}

In particular, Idempotents are split in categories of modules, comodules,
crossed modules over usual Hopf algebra.

\begin{proposition}\label{Radford}
Let $\cal C$ be a braided category with split idempotents,
$A \matrix{i_A\cr \longrightarrow\cr \longleftarrow\cr p_A} H$
a Hopf algebra progection in $\cal C$ (i.e. $i_A$ and $p_A$ be Hopf algebra
morphisms with $p_A\circ i_A={\rm id}_A$).
Then there exists a Hopf algebra $B$ living in braided category ${\cal C}_A^A$
such that $H\simeq A\ltimes B$.
\end{proposition}

\begin{proof}
Define the following idempotent
\begin{equation}
\Pi:=\cdot_H\circ (i_A\circ S_A\circ p_A\otimes{\rm id})\circ\Delta_H:\,
     H\rightarrow H
\end{equation}
It's convenient to construct Hopf algebra $B$ in two steps.
Primarily, consider $H$ with new 'multiplication', 'comultiplication',
'antipode', right 'action' and 'coaction' of $A$:
$$\widehat\cdot :=\Pi\circ\cdot_H\circ (\Pi\otimes\Pi )=
\Pi\circ\cdot_H\circ ({\rm id}\otimes\Pi )=
\cdot_H\circ (\Pi\otimes\Pi )$$
$$\widehat\Delta :=(\Pi\otimes\Pi)\circ\Delta_H\circ\Pi=
(\Pi\otimes\Pi)\circ\Delta_H=
({\rm id}\otimes\Pi)\circ\Delta_H\circ\Pi$$
$$\widehat\triangleleft :=\Pi\circ\cdot_H\circ (\Pi\otimes i_A)=
\Pi\circ\cdot_H\circ ({\rm id}\otimes i_A)=
{}_{\rm ad}\triangleleft\circ (\Pi\otimes i_A)$$
$$\widehat\beta :=(\Pi\otimes p_A)\circ\Delta_H\circ\Pi=
({\rm id}\otimes p_A)\circ\Delta_H\circ\Pi=
(\Pi\otimes p_A)\circ{}^{\rm ad}\beta$$
$$\widehat S:=
\widehat\triangleleft\circ (S_H\otimes{\rm id}_A)\circ\widehat\beta=
\Pi\circ\cdot_H\circ (S_H\otimes i_A\circ p_A)\circ\Delta_H\circ\Pi=
\cdot_H\circ (S_H\otimes i_A\circ p_A)\circ\Delta_H$$
and with old unit and counit.
All axioms are true as $H$ whould be a Hopf algebra in category of crossed
modules over $A$ with the following exceptions.
Result of 'multiplication' of unit or 'action' of unit is equal to projection
$\Pi$ but not identity, and similarly for counit.

Second step: let
$H \matrix{p_B\cr \longrightarrow\cr \longleftarrow\cr i_B} B$
split $\Pi$ in $\cal C$, i.e.
$p_B\circ i_B=\Pi$, $i_B\circ p_B={\rm id}_B$.
Define on $B$ the following multiplication, unit, comultiplication, counit,
antipode, rigth action and coaction of $A$:
$$\cdot_B:=p_B\circ\cdot_H\circ (i_B\otimes i_B)\qquad
  \eta_B:=\eta_H\circ i_B$$
$$\Delta_B:=(p_B\otimes p_B)\circ\Delta_H\circ i_B\qquad
  \epsilon_B:=p_B\circ\epsilon_H$$
$$\triangleleft:=p_B\circ\cdot_H\circ (i_B\otimes i_A)\qquad
  \beta:=(p_B\otimes p_A)\circ\Delta_H\circ i_B$$
$$S_B:=
p_B\circ\cdot_H\circ (S_H\otimes i_A\circ p_A)\circ\Delta_H\circ i_B$$
$B$ is a Hopf algebra in category ${\cal C}^A_A$.
This fact can be obtaned directly from previous result about $H$ with new
structure.

Notice also that $i_B$ is equalizer of first and $p_B$ is coequalizer of
second pair of maps
$$H \matrix{p_A\otimes{\rm id})\circ\Delta_H\cr
	    \longrightarrow\cr \longrightarrow\cr
            \eta _A\otimes{\rm id}} A\otimes H\qquad
  A\otimes H
    \matrix{\cdot_H\circ (i_A\otimes{\rm id})\cr
	    \longrightarrow\cr \longrightarrow\cr
            \epsilon _A\otimes{\rm id}}                 H$$
$\theta :=\cdot_H\circ (i_A\otimes i_B) :\, A\ltimes B\rightarrow H$
is Hopf algera isomorphism with inverse
$\theta_{-1}:=(p_A\otimes p_B)\circ\Delta_H$
\end{proof}

Let $A$ be a Hopf algebra in category $\cal C$, $B$ a Hopf algebra in
${\cal C}^A_A$ and $X$ be a right module over $B$. The last means that $X$ is
crossed module over $A$ and action of $B$ is crossed module morphism.
It's easy verify that formula
\begin{equation}
\matrix{\object{X}\Step\object{A\ltimes B}\cr
	\vvbox{\hbox{\Ru}}}
\quad :=\quad
\matrix{\object{X}\step\object{A}\step\object{B}\cr
	\vvbox{\hbox{\ru\step\id}
	       \hbox{\Ru}}}
\end{equation}
define ($A\!\ltimes\! B$)-module structure on $X$.
Moreover in this way one can construct full imbedding of categories
$$({\cal C}^A_A)_B\hookrightarrow{\cal C}_{A\ltimes B}$$

\begin{proposition}
Let $A$ be a Hopf algebra in category $\cal C$ and $B$ a Hopf algebra in
category ${\cal C}^A_A$. Then braided categories $({\cal C}^A_A)^B_B$ and
${\cal C}^{A\ltimes B}_{A\ltimes B}$ are isomorphic.
\end{proposition}

\begin{proof}
($A\!\ltimes\! B$)-modile structure on object $X$ of $({\cal C}_A^A)^B_B$
is defined above and $A\ltimes B$-comodule structure is obtained using
input-output symmetry.
Nontrivial part is to verify crossed module axiom for $X$
(see figures \ref{proofovercross1} - \ref{proofovercross2}).

Conversely, $A$-, ($B$)- crossed module structure on objects of
${\cal C}^{A\ltimes B}_{A\ltimes B}$ is defined by means of $i_A$, $p_A$
($i_B$, $p_B$) (see proof of the proposition \ref{Radford}).
\end{proof}

So for object $C$ in $\cal C$ to be a Hopf algebra in
${\cal C}^{A\ltimes B}_{A\ltimes B}$ or in $({\cal C}^A_A)^B_B$
are eqivalent. In this case on $A\otimes B\otimes C$ there exist two Hopf
algebra structures $(A\ltimes B)\ltimes C$ and $A\ltimes (B\ltimes C)$.

\begin{proposition} {\em (Transitivity of cross product)}
Hopf algebras $(A\ltimes B)\ltimes C$ and $A\ltimes (B\ltimes C)$ coincide.
\label{transitivity}
\end{proposition}

\newfigure
\begin{figure}
\begin{displaymath}
\matrix{\object{X}\step\object{Y}\Step\object{H}\Step\cr
\vvbox{\hbox{\id\step\id\step\cd}
       \hbox{\id\step\hx\step\cd}
       \hbox{\hx\step\hx\Step\id}
       \hbox{\id\step\ru\step\Ru}
       \hhbox{\id\step\rd\step\Rd}
       \hbox{\hx\step\hx\Step\id}
       \hbox{\id\step\hx\step\cu}
       \hbox{\id\step\id\step\cu}}\cr
\object{X}\step\object{Y}\Step\object{H}\Step}
\quad ={}\quad
\vvbox{\hhbox{\id\step\hstep\id\step\cd}
       \hbox{\id\step\hstep\hx\step\id}
       \hhbox{\id\step\cd\hstep\id\step\id}
       \hbox{\hx\step\id\hstep\ru}
       \hbox{\id\step\ru\hstep\id}
       \hbox{\id\step\rd\hstep\id}
       \hbox{\hx\step\id\hstep\rd}
       \hhbox{\id\step\cu\hstep\id\step\id}
       \hbox{\id\step\hstep\hx\step\id}
       \hhbox{\id\step\hstep\id\step\cu}}
\quad ={}\quad
\vvbox{\hhbox{\id\Step\id\step\cd}
       \hbox{\id\Step\hx\step\id}
       \hhbox{\hrd\step\cd\hstep\ru}
       \hbox{\id\hstep\hx\step\id\hstep\id}
       \hhbox{\hru\step\cu\hstep\rd}
       \hbox{\id\Step\hx\step\id}
       \hhbox{\id\Step\id\step\cu}}
\quad ={}
\end{displaymath}
\begin{quotation}
We use coherence and associativty (first, third, fifth equalities),
crossed module axiom (second and fourth ones).
\end{quotation}
\caption{Crossed module axiom for tensor product (part 1).}
\label{prooftensorproduct1}
\end{figure}
\newfigure
\begin{figure}
\begin{displaymath}
 ={}\quad
\vvbox{\hbox{\id\step\step\id\step\cd}
       \hbox{\id\step\step\hx\step\cd}
       \hbox{\id\step\step\id\step\hx\Step\id}
       \hhbox{\rd\step\id\step\id\step\Ru}
       \hbox{\id\step\hx\step\id\step\id}
       \hhbox{\ru\step\id\step\id\step\Rd}
       \hbox{\id\step\step\id\step\hx\Step\id}
       \hbox{\id\step\step\hx\step\cu}
       \hbox{\id\step\step\id\step\cu}}
\quad ={}\quad
\vvbox{\hbox{\id\step\step\id\step\cd}
       \hbox{\id\step\step\hx\step\hstep\hcd}
       \hhbox{\rd\step\id\step\hrd\step\id\step\id}
       \hbox{\id\step\hx\step\id\hstep\hx\step\id}
       \hhbox{\ru\step\id\step\hru\step\id\step\id}
       \hbox{\id\step\step\hx\step\hstep\hcu}
       \hbox{\id\step\step\id\step\cu}}
\quad ={}\quad
\vvbox{\hbox{\rd\step\rd\step\id}
       \hbox{\id\step\hx\step\id\hstep\hcd}
       \hhbox{\id\step\id\step\cu\hstep\id\step\id}
       \hbox{\id\step\id\step\hstep\hx\step\id}
       \hhbox{\id\step\id\step\cd\hstep\id\step\id}
       \hbox{\id\step\hx\step\id\hstep\hcu}
       \hbox{\ru\step\ru\step\id}}
\end{displaymath}
\caption{Crossed module axiom for tensor product (part 2).}
\label{prooftensorproduct2}
\end{figure}

\newfigure
\begin{figure}
\begin{displaymath}
\matrix{\object{X}\step\object{Y}\step\hstep\object{H}\hstep\cr
\vvbox{\hhbox{\id\step\id\step\cd}
       \hbox{\id\step\hx\step\id}
       \hbox{\ru\step\ru}
       \hhbox{\id\Step\rd}
       \hbox{\x\step\id}
       \hbox{\id\Step\ru}}\cr
	\object{X}\Step\object{Y}\step}
\quad ={}\quad
\vvbox{\hhbox{\id\step\id\step\cd}
       \hbox{\id\step\hx\step\id}
       \hhbox{\id\step\id\step\ru}
       \hbox{\id\step\id\step\rd}
       \hbox{\id\step\hx\step\id}
       \hbox{\hx\step\hcu}
       \hhbox{\id\step\d\step\id}
       \hbox{\id\step\hstep\ru}}
\quad ={}\quad
\vvbox{\hhbox{\id\step\hrd\step\cd}
       \hbox{\id\step\id\hstep\hx\step\id}
       \hhbox{\id\step\hru\step\cu}
       \hbox{\hx\Step\id}
       \hbox{\id\step\Ru}}
\quad ={}\quad
\vvbox{\hbox{\id\step\rd\hstep\id}
       \hbox{\hx\step\id\hstep\id}
       \hbox{\id\step\ru\hstep\id}
       \hhbox{\id\step\id\step\cd}
       \hbox{\id\step\hx\step\id}
       \hbox{\ru\step\ru}}
\end{displaymath}
\begin{quotation}
First, second and third equalities follow from module, crossed module and
and comodule axioms respectively.

Input-output symmetrical diagrams show that braiding is comodule map.
\end{quotation}
\caption{Braiding in ${\cal C}_H^H$ is module map.}
\label{proofmodulemap}
\end{figure}

\newfigure
\begin{figure}
\begin{displaymath}
{}^{({\cal C}^H_H)}\Psi_{X\otimes Y,Z}
:=
\matrix{\object{X}\step\object{Y}\step\hstep\object{Z}\step\hstep\cr
        \vvbox{\hbox{\id\step\id\step\hstep\rd}
               \hhbox{\id\step\id\step\dd\step\id}
               \hbox{\id\step\hx\step\hstep\id}
               \hbox{\hx\step\id\step\hcd}
               \hbox{\id\step\id\step\hx\step\id}
               \hbox{\id\step\ru\step\ru}}}
=\enspace
\vvbox{\hbox{\id\step\id\step\Rd}
       \hhbox{\id\step\id\step\rd\step\id}
       \hbox{\id\step\hx\step\id\step\id}
       \hbox{\hx\step\hx\step\id}
       \hbox{\id\step\ru\step\ru}}
\enspace =\enspace
\vvbox{\hbox{\id\Step\id\step\rd}
       \hbox{\id\Step\hx\step\id}
       \hbox{\id\step\dd\step\ru}
       \hbox{\id\step\rd\step\id}
       \hbox{\hx\step\id\step\id}
       \hbox{\id\step\ru\step\id}}
=:
{}^{({\cal C}^H_H)}\Psi_{X,Z}\circ{}^{({\cal C}^H_H)}\Psi_{Y,Z}
\end{displaymath}
\begin{quotation}
We use comodule axiom and coherence.
To prove second hexagon one needs input-output symmetrical diagrams.
Crossed module axiom is not required.
\end{quotation}
\caption{Hexagon identities in ${\cal C}_H^H$.}
\end{figure}

\newfigure
\begin{figure}
\begin{displaymath}
{}^{({\cal C}_H^H)}\Psi_{X,Y}^{-1}\circ{}^{({\cal C}_H^H)}\Psi_{X,Y}
\quad :=\quad
\matrix{\object{X}\Step\object{Y}\step\cr
\vvbox{\hbox{\id\Step\rd}
       \hbox{\x\step\id}
       \hbox{\rd\step\ru}
       \hbox{\id\step\hxx}
       \hbox{\hxx\step\SS}
       \hbox{\id\step\hxx}
       \hbox{\ru\step\id}}}
\quad ={}\quad
\vvbox{\hbox{\id\step\rd}
       \hbox{\id\step\hxx}
       \hbox{\ru\step\rd}
       \hbox{\id\Step\id\step\SS}
       \hbox{\id\Step\hxx}
       \hbox{\Ru\step\id}}
\quad ={}\quad
\vvbox{\hbox{\id\step\rd}
       \hhbox{\id\step\id\hstep\cd}
       \hbox{\id\step\id\hstep\hxx}
       \hbox{\id\step\id\hstep\id\step\SS}
       \hhbox{\id\step\id\hstep\cu}
       \hbox{\id\step\hxx}
       \hbox{\ru\step\id}}
\quad ={}\quad
{\rm id}_{X,Y}
\end{displaymath}
\begin{quotation}
Proof of $\;{}^{({\cal C}_H^H)}\Psi\circ{}^{({\cal C}_H^H)}\Psi^-1={\rm id}\;$
use input-output symmetrical diagrams.
\end{quotation}
\caption{Braiding in ${\cal C}_H^H$ is invertible.}
\label{proofinverse}
\end{figure}

\newfigure
\begin{figure}
\begin{displaymath}
\matrix{\object{X^\vee}\Step\hstep\object{H}\hstep\cr
        \vvbox{\hbox{\rd\step\hcd}
               \hbox{\id\step\hx\step\id}
	       \hbox{\ru\step\hcu}}\cr
       \object{X^\vee}\Step\hstep\object{H}\hstep}
\quad :=\quad
\vvbox{\hhbox{\Step\hcoev\Step\id\step\hstep\id}
       \hhbox{\Step\id\step\rd\step\id\step\cd}
       \hbox{\Step\id\step\hx\step\id\step\id\step\id}
       \hbox{\Step\id\step\SS\step\hev\dd\step\id}
       \hhbox{\Step\id\step\d\step\dd\Step\id}
       \hbox{\Step\id\step\hstep\hx\Step\hstep\id}
       \hhbox{\Step\id\step\dd\step\d\Step\id}
       \hbox{\hcoev\step\id\step\S\Step\d\step\id}
       \hbox{\id\step\id\step\hx\Step\step\id\step\id}
       \hhbox{\id\step\ru\step\id\Step\step\cu}
       \hbox{\id\step\ev\Step\step\hstep\id}}
\quad ={}\quad
\vvbox{\hhbox{\Step\id\step\cd}
       \hbox{\Step\id\step\S\step\id}
       \hbox{\hcoev\step\hx\step\S}
       \hbox{\id\step\ru\step\id\step\id}
       \hhbox{\id\step\rd\step\id\step\id}
       \hbox{\id\step\id\step\hx\step\SS}
       \hbox{\id\step\hev\step\SS\step\id}
       \hhbox{\id\Step\step\cu}}
\quad ={}\quad
\vvbox{\hbox{\Step\id\step\hstep\S}
       \hhbox{\Step\id\step\cd}
       \hbox{\Step\id\step\hxx}
       \hbox{\hcoev\step\hx\step\id}
       \hbox{\id\step\ru\step\id\step\id}
       \hhbox{\id\step\rd\step\id\step\id}
       \hbox{\id\step\id\step\hx\step\id}
       \hbox{\id\step\hev\step\hx}
       \hhbox{\id\Step\step\cu}
       \hbox{\id\Step\step\hstep\SS}}
\quad ={}
\end{displaymath}
\begin{quotation}
The second, 4th and 7th equalities use coherence
(to the third diagram we also add identical composition $S^{-1}\circ S$).
Antipode is algebra and coalgebra antihomomorphism.
That imply the the third equality.
The 5th is crossed module axiom.
Steps need to prove the 6th are analogous ones
used in the second--4th equalities.
\end{quotation}
\caption{Crossed module axiom for right dual (part 1).}
\label{proofdual1}
\end{figure}

\newfigure
\begin{figure}
\begin{displaymath}
 ={}\quad
\vvbox{\hbox{\Step\hstep\id\step\S}
       \hbox{\hcoev\step\hstep\hx}
       \hhbox{\id\step\id\step\cd\hstep\id}
       \hbox{\id\step\hx\step\id\hstep\id}
       \hbox{\id\step\id\ru\step\id\hstep\id}
       \hhbox{\id\step\id\rd\step\id\hstep\id}
       \hbox{\id\step\hx\step\id\hstep\id}
       \hhbox{\id\step\id\step\cu\hstep\id}
       \hbox{\id\step\id\step\hstep\hx}
       \hhbox{\id\step\id\step\dd\step\id}
       \hbox{\id\step\hev\step\hstep\SS}}
\quad ={}\quad
\vvbox{\hbox{\Step\step\id\step\S}
       \hbox{\hcoev\Step\hx}
       \hhbox{\id\step\hrd\step\cd\hstep\id}
       \hbox{\id\step\id\hstep\hx\step\id\hstep\id}
       \hhbox{\id\step\hru\step\cu\hstep\id}
       \hbox{\id\step\id\Step\hx}
       \hbox{\id\step\ev\step\SS}}
\quad ={}\quad
\vvbox{\hbox{\Step\step\id\step\cd}
       \hbox{\Step\step\hx\Step\id}
       \hbox{\Step\dd\step\d\step\S}
       \hbox{\step\dd\step\hcoev\step\hx}
       \hbox{\dd\Step\id\step\ru\step\id}
       \hbox{\id\step\hcoev\step\id\step\ev}
       \hbox{\id\step\id\step\rd\d}
       \hbox{\id\step\id\step\hx\step\id}
       \hbox{\hx\step\S\step\hev}}
\quad =:\quad
\matrix{\object{X^\vee}\hstep\step\object{H}\hstep\cr
        \vvbox{\hhbox{\id\step\cd}
               \hbox{\hx\step\id}
               \hhbox{\id\step\ru}
               \hbox{\id\step\rd}
               \hbox{\hx\step\id}
               \hhbox{\id\step\cu}}\cr
         \object{X^\vee}\hstep\step\object{H}\hstep}
\end{displaymath}
\caption{Crossed module axiom for right dual (part 2).}
\label{proofdual2}
\end{figure}

\newfigure
\begin{figure}
\begin{displaymath}
\matrix{\hstep\object{H}\step\hstep\object{X}\cr
        \vvbox{\hhbox{\cd\step\id}
               \hbox{\SS\step\hx}
               \hbox{\hxx\step\id}
               \hbox{\ru\step\id}
               \hhbox{\rd\step\id}
               \hbox{\hx\step\id}
               \hbox{\S\step\hx}
               \hhbox{\cu\step\id}}\cr
        \hstep\object{H}\step\hstep\object{X}}
\quad ={}\quad
\vvbox{\hbox{\SS\step\id}
       \hbox{\hxx}
       \hhbox{\id\step\d}
       \hhbox{\id\step\cd}
       \hbox{\hx\step\id}
       \hhbox{\id\step\ru}
       \hbox{\id\step\rd}
       \hbox{\hx\step\id}
       \hhbox{\id\step\cu}
       \hhbox{\id\step\dd}
       \hbox{\hx}
       \hbox{\S\step\id}}
\quad ={}\quad
\vvbox{\hbox{\SS\Step\id}
       \hbox{\xx}
       \hhbox{\hrd\step\cd}
       \hbox{\id\hstep\hx\step\id}
       \hhbox{\hru\step\cu}
       \hbox{\x}
       \hbox{\S\Step\id}}
\quad ={}\quad
\vvbox{\hbox{\hstep\id\step\hstep\rd}
       \hbox{\hstep\id\step\hstep\hx}
       \hbox{\hcd\step\S\step\id}
       \hbox{\id\step\hx\step\id}
       \hbox{\hcu\step\SS\step\id}
       \hbox{\hstep\id\step\hstep\hxx}
       \hbox{\hstep\id\step\hstep\ru}}
\end{displaymath}
\begin{quotation}
The first and third equalities mean that antipode is algebra and coalgerba
antihomomorphism.
The second is crossed module axiom.
\end{quotation}
\caption{Crossed module axiom for $X^S$.}
\label{proofleftright}
\end{figure}

\newfigure
\thispagestyle{empty}
\addtocounter{page}{-1}
\begin{figure}
\begin{displaymath}
\vvbox{\hbox{\rd}
       \hbox{\id\step\S}
       \hbox{\ru}
       \hhbox{\rd}
       \hbox{\hxx}
       \hbox{\SS\step\id}}
\quad ={}\quad
\vvbox{\hbox{\Rd}
       \hbox{\id\Step\S}
       \hbox{\id\step\cd}
       \hbox{\hx\step\cd}
       \hbox{\id\step\hx\Step\id}
       \hbox{\S\step\id\step\Ru}
       \hbox{\id\step\id\step\Rd}
       \hbox{\id\step\hx\Step\id}
       \hbox{\hx\step\cu}
       \hbox{\id\step\cu}
       \hbox{\xx}
       \hbox{\SS\Step\id}}
\quad ={}\quad
\vvbox{\hbox{\Rd}
       \hbox{\id\Step\S}
       \hbox{\id\step\cd}
       \hbox{\hx\Step\id}
       \hhbox{\id\step\hrd\step\cd}
       \hbox{\S\step\id\hstep\hx\step\id}
       \hhbox{\id\step\hru\step\cu}
       \hbox{\id\step\xx}
       \hhbox{\cu\Step\id}
       \hbox{\hstep\SS\hstep\Step\id}}
\quad ={}\quad
\vvbox{\hbox{\rd}
       \hbox{\hx}
       \hbox{\id\step\Rd}
       \hbox{\id\step\rd\step\S}
       \hbox{\id\step\id\step\hx}
       \hbox{\S\step\ru\step\id}
       \hhbox{\id\step\id\step\hstep\cd}
       \hbox{\id\step\id\step\hstep\SS\step\id}
       \hbox{\id\step\id\step\hstep\hxx}
       \hhbox{\id\step\id\step\hstep\cu}
       \hbox{\id\step\x}
       \hbox{\hxx\Step\id}
       \hhbox{\cu\Step\id}}
\quad ={}\quad
\vvbox{\hbox{\rd}
       \hbox{\hx}
       \hbox{\S\step\rd}
       \hbox{\id\step\id\step\S}
       \hbox{\id\step\ru}}
\end{displaymath}
\begin{quotation}
The first equality follow from associativity and antipode axiom.
The second is crossed module axiom.
In the third we use that antipode is algebra and coalgebra antihomomorphism.
The 4th is antipode axiom again.

To prove that $X^S\buildrel{S^2}\over\rightarrow{}^SX$ is module map one needs
input-output symmetrical diagramms.
\end{quotation}
\caption{$X^S\buildrel{S^2}\over\rightarrow{}^SX$ is comodule map.}
\label{proofantipodemodulemap}
\end{figure}

\newfigure
\begin{figure}
\begin{displaymath}
\matrix{\object{X}\cr
        \vvbox{\hbox{\begin{picture}(0,2)
                     \put(0,0){\line(0,1){0.4}}
                     \put(0,1.6){\line(0,1){0.4}}\put(0,1){\circle{1.2}}
                     \put(-0.6,0.4){\makebox(1.2,1.2)[cc]{$\scriptstyle S^2$}}
                     \end{picture}}
               \hbox{\begin{picture}(0,2)
                     \put(0,0){\line(0,1){0.4}}
                     \put(0,1.6){\line(0,1){0.4}}\put(0,1){\circle{1.2}}
                     \put(-.6,.4){\makebox(1.2,1.2)[cc]{$\scriptstyle S^{-2}$}}
                     \end{picture}}}\cr
	\object{X}}
\quad :=\quad
\vvbox{\hbox{\rd}
       \hbox{\id\step\S}
       \hhbox{\ru}
       \hbox{\rd}
       \hbox{\hxx}
       \hbox{\SS\step\id}
       \hbox{\SS\step\id}
       \hbox{\hxx}
       \hbox{\ru}}
\quad ={}\quad
\vvbox{\hbox{\rd}
       \hbox{\hx}
       \hbox{\S\step\rd}
       \hbox{\id\step\id\step\S}
       \hbox{\SS\step\ru}
       \hbox{\hxx}
       \hbox{\ru}}
\quad ={}\quad
\vvbox{\hbox{\Rd}
       \hhbox{\id\step\hstep\cd}
       \hbox{\id\step\hstep\S\step\id}
       \hhbox{\id\step\hstep\cu}
       \hbox{\Ru}}
\quad ={}\quad
\vvbox{\hbox{\id}
       \hbox{\id}}
\end{displaymath}
\begin{quotation}
In the second equality we use identity has proved
in figure ref{proofantipodemodulemap}.
\end{quotation}
\caption{Map $X^S\buildrel{S^2}\over\rightarrow{}^SX$ is invertible.}
\label{proofantipodeinvert}
\end{figure}

\newfigure
\thispagestyle{empty}
\addtocounter{page}{-1}
\begin{figure}
\begin{displaymath}
\vvbox{\hbox{\id\Step\rd}
       \hbox{\x\step\id}
       \hhbox{\id\Step\ru}
       \hbox{\rd\step\rd}
       \hbox{\id\step\hx\step\id}
       \hhbox{\id\step\id\step\cu}
       \hbox{\id\step\id\step\hstep\S}
       \hhbox{\id\step\id\step\cd}
       \hbox{\id\step\hx\step\id}
       \hbox{\ru\step\ru}
       \hhbox{\id\Step\rd}
       \hbox{\x\step\id}
       \hbox{\id\Step\ru}}
\quad=\quad
\vvbox{\hbox{\rd\Step\Rd}
       \hbox{\id\step\x\Step\id}
       \hbox{\id\step\rd\step\x}
       \hbox{\hx\step\id\step\S\Step\S}
       \hbox{\id\step\ru\cd\step\hdcd}
       \hbox{\id\step\hx\Step\hx\step\id}
       \hbox{\ru\step\x\step\id\step\id}
       \hbox{\Ru\Step\ru\step\id}
       \hbox{\d\Step\step\Ru}
       \hbox{\step\d\Step\rd}
       \hbox{\Step\x\step\id}
       \hbox{\Step\id\Step\ru}}
\quad=\quad
\vvbox{\hbox{\x}
       \hbox{\rd\step\rd}
       \hbox{\id\step\S\step\id\step\S}
       \hbox{\ru\step\id\step\hdcd}
       \hbox{\d\step\hx\step\id}
       \hbox{\step\ru\step\ru}
       \hbox{\step\id\Step\rd}
       \hbox{\step\x\step\id}
       \hbox{\step\id\Step\ru}}
\quad=\quad
\vvbox{\hbox{\x}
       \hbox{\rd\step\rd}
       \hbox{\id\step\S\step\id\step\S}
       \hbox{\ru\step\ru}
       \hbox{\x}}
\end{displaymath}
\begin{quotation}
The first equality use that ${}^{({\cal C}_H^H)}\Psi$ is comodule
morphism, antipode is antialgebra morphism, bialgebra axiom, module axiom.
The second use that antipode is anticoalgebra morphism and antipode axiom.
The third use that ${}^{({\cal C}_H^H)}\Psi$ is module morphism,
antipode is anticoalgebra morphism and antipode axiom.
\end{quotation}
\caption{${}^{({\cal C}_H^H)}\Psi\circ S^2\circ{}^{({\cal C}_H^H)}\Psi=
          {}^{\cal C}\Psi\circ (S^2\otimes S^2)\circ{}^{\cal C}\Psi$}
\label{proofantipodetensor}
\end{figure}

\newfigure
\thispagestyle{empty}
\addtocounter{page}{-1}
\begin{figure}
\vskip -1 true cm
\begin{displaymath}
\matrix{\object{X_{[{\rm op}]}}\hstep\step\object{H}\step\cr
        \vvbox{\hhbox{\id\step\cd}
               \hbox{\hx\step\id}
               \hhbox{\id\step\ru}
               \hbox{\id\step\rd}
               \hbox{\hx\step\id}
               \hhbox{\id\step\cu}}\cr
       \hbox{(in category}\enspace\overline{\cal C})}
\quad ={}\quad
\matrix{\object{X}\step\hstep\object{H}\hstep\cr
	\vvbox{\hhbox{\id\step\cd}
	       \hbox{\id\step\hxx}
	       \hbox{\hxx\step\id}
	       \hbox{\id\step\ru}
	       \hhbox{\id\step\rd}
	       \hbox{\id\step\hxx}
	       \hbox{\id\step\SS\step\id}
	       \hbox{\id\step\hxx}
	       \hbox{\hxx\step\id}
	       \hhbox{\id\step\cu}}}
\quad ={}\qquad
\vvbox{\hbox{\hstep\rd\step\hcd}
       \hbox{\hstep\hxx\step\hxx}
       \hbox{\hstep\id\step\hxx\step\id}
       \hhbox{\cd\hstep\id\step\ru}
       \hbox{\hxx\hstep\id\step\rd}
       \hbox{\id\step\id\hstep\id\step\hxx}
       \hbox{\SS\step\id\hstep\id\step\SS\step\id}
       \hhbox{\cu\hstep\id\step\id\step\id}
       \hbox{\hstep\id\step\id\step\hxx}
       \hbox{\hstep\id\step\hxx\step\id}
       \hbox{\hstep\hxx\step\id\step\id}
       \hhbox{\hstep\id\step\d\hstep\cu}
       \hhbox{\hstep\id\step\hstep\cu}}
\quad ={}\quad
\vvbox{\hbox{\rd\Step\id}
       \hbox{\hxx\Step\id}
       \hhbox{\id\step\hrd\step\cd}
       \hbox{\SS\step\id\hstep\hx\step\id}
       \hhbox{\id\step\hru\step\cu}
       \hbox{\id\step\xx}
       \hbox{\id\step\id\Step\rd}
       \hbox{\id\step\id\Step\hxx}
       \hbox{\id\step\id\Step\SS\step\id}
       \hbox{\id\step\id\Step\hxx}
       \hbox{\id\step\xx\step\id}
       \hbox{\hxx\Step\id\step\id}
       \hhbox{\id\step\d\hstep\step\cu}
       \hbox{\id\step\hstep\cu}}
\quad ={}
\end{displaymath}
\begin{quotation}
In the second and 6th equalities we use antipode axiom.
The third and 5th follow from associativity and coherence.
The 4th is crossed module axiom.
\end{quotation}
\caption{Crossed module axiom for $X_{[{\rm op}]}$ (part 1).}
\label{proofcop1}
\end{figure}

\newfigure
\begin{figure}
\begin{displaymath}
 ={}\quad
\vvbox{\hbox{\rd\step\hstep\id}
       \hbox{\hxx\step\hcd}
       \hbox{\SS\step\hx\step\id}
       \hbox{\id\step\id\step\ru}
       \hhbox{\id\step\id\step\rd}
       \hbox{\id\step\hx\step\id}
       \hhbox{\id\step\id\step\cu}
       \hhbox{\id\step\id\step\cd}
       \hbox{\id\step\id\step\SS\step\id}
       \hbox{\id\step\id\step\hxx}
       \hbox{\id\step\hxx\step\id}
       \hbox{\hxx\step\hcu}
       \hhbox{\id\step\d\step\id}
       \hhbox{\id\step\hstep\cu}}
\quad ={}\quad
\vvbox{\hbox{\step\rd\step\hstep\id}
       \hbox{\step\hxx\step\hcd}
       \hbox{\dd\step\hx\step\id}
       \hbox{\SS\step\dd\step\ru}
       \hhbox{\id\step\id\Step\rd}
       \hhbox{\id\step\id\Step\id\hstep\cd}
       \hbox{\id\step\id\Step\id\hstep\hxx}
       \hbox{\id\step\id\Step\id\hstep\id\step\SS}
       \hhbox{\id\step\id\Step\id\hstep\cu}
       \hbox{\id\step\id\Step\hxx}
       \hbox{\id\step\cu\step\id}
       \hbox{\cu\step\dd}
       \hhbox{\step\d\step\dd}
       \hbox{\step\hstep\hxx}}
\quad ={}\quad
\vvbox{\hbox{\rd\step\hstep\id}
       \hbox{\hxx\step\hstep\id}
       \hbox{\SS\step\id\step\hcd}
       \hbox{\hxx\step\hxx}
       \hbox{\id\step\hxx\step\id}
       \hbox{\ru\step\hcu}}
\quad :=\quad
\matrix{\object{X_{[{\rm op}]}}\Step\hstep\object{H}\step\cr
        \vvbox{\hbox{\rd\step\hcd}
               \hbox{\id\step\hx\step\id}
	       \hbox{\ru\step\hcu}}\cr
        \hbox{(in category}\enspace\overline{\cal C})}
\end{displaymath}
\caption{Crossed module axiom for $X_{[{\rm op}]}$ (part 2).}
\label{proofcop2}
\end{figure}

\newfigure
\begin{figure}
\begin{displaymath}
\matrix{\object{{}^{[\vee ]}X}\Step\hstep\object{{}^\vee H}\hstep\cr
        \vvbox{\hbox{\rd\step\hcd}
               \hbox{\id\step\hx\step\id}
	       \hbox{\ru\step\hcu}}\cr
       \object{{}^{[\vee]}X}\Step\hstep\object{{}^\vee H}\hstep}
\quad :=\quad
\matrix{\object{X}\Step\hstep\object{H}\Step\Step\step\cr
\vvbox{\hbox{\id\Step\hstep\id\hstep\coev}
       \hbox{\id\Step\hstep\id\hstep\id\step\hcoev\d}
       \hhbox{\id\Step\hstep\id\hstep\cu\hstep\dd\step\id}
       \hhbox{\id\step\coev\hstep\ev\hstep\dd\step\hstep\id}
       \hbox{\ru\step\id\step\dd\Step\id}
       \hbox{\d\step\hx\Step\dd\hstep\hcoev}
       \hbox{\step\hx\step\d\step\id\step\hcd\hstep\id}
       \hbox{\step\id\step\rd\step\d\hev\step\id\hstep\id}
       \hbox{\step\hx\step\id\Step\ev\hstep\id}
       \hhbox{\step\id\step\cu\Step\Step\hstep\id}}}
\quad ={}\quad
\vvbox{\hbox{\hx\step\hstep\coev}
       \hhbox{\id\step\id\step\cd\step\hstep\id}
       \hbox{\id\step\hx\step\id\step\hstep\id}
       \hbox{\id\step\id\step\ru\step\hstep\id}
       \hhbox{\id\step\id\step\rd\step\hstep\id}
       \hbox{\id\step\hx\step\id\step\hstep\id}
       \hbox{\hx\step\hcu\step\hstep\id}
       \hhbox{\id\step\d\step\id\Step\id}
       \hbox{\id\step\hstep\hev\Step\id}}
\quad ={}\quad
\end{displaymath}
\begin{quotation}
The second and 4th equalities use coherance.
The third is crossed module axiom.
\end{quotation}
\caption{Crossed module axiom for ${}^{[\vee ]}X$ (part 1).}
\label{proofvee1}
\end{figure}

\newfigure
\thispagestyle{empty}
\addtocounter{page}{-1}
\begin{figure}
\begin{displaymath}
\quad ={}\quad
\vvbox{\hbox{\hxx\Step\hcoev}
       \hhbox{\id\step\hrd\step\cd\hstep\id}
       \hbox{\id\step\id\hstep\hx\step\id\hstep\id}
       \hhbox{\id\step\hru\step\cu\hstep\id}
       \hbox{\hx\Step\id\step\id}
       \hbox{\id\step\ev\step\id}}
\quad ={}\quad
\vvbox{\hbox{\id\step\id\hstep\coev}
       \hhbox{\id\step\id\hstep\id\Step\d}
       \hhbox{\id\step\id\hstep\id\step\coev\hstep\id}
       \hhbox{\id\step\id\hstep\cu\hstep\dd\hstep\id}
       \hbox{\d\hev\dd\step\id}
       \hbox{\step\hx\step\dd}
       \hbox{\step\id\step\hx}
       \hbox{\step\id\step\id\step\rd}
       \hbox{\step\id\step\hx\step\id}
       \hhbox{\step\id\step\id\step\ev}
       \hhbox{\step\id\step\id\step\coev\hstep\coev}
       \hbox{\step\id\step\ru\dd\hcd\hstep\id}
       \hbox{\step\hx\step\hev\dd\hstep\id}
       \hbox{\step\id\step\ev\step\hstep\id}}
\quad ={}\quad
\matrix{\object{{}^{[\vee]}X}\hstep\step\object{{}^\vee H}\step\cr
        \vvbox{\hhbox{\id\step\cd}
               \hbox{\hx\step\id}
               \hhbox{\id\step\ru}
               \hbox{\id\step\rd}
               \hbox{\hx\step\id}
               \hhbox{\id\step\cu}}\cr
       \object{{}^{[\vee]}X}\hstep\step\object{{}^\vee H}\step}
\end{displaymath}
\caption{Crossed module axiom for ${}^{[\vee ]}X$ (part 2).}
\label{proofvee2}
\end{figure}

\newfigure
\begin{figure}
\begin{displaymath}
\matrix{\object{A\!\ltimes\! B}\Step\object{A\!\ltimes\! B}\cr
        \vvbox{\hbox{\cu}
               \hbox{\cd}}\cr
        \object{A\!\ltimes\! B}\Step\object{A\!\ltimes\! B}}
\quad :=\quad
\matrix{\object{A}\Step\object{B}\step\hstep\object{A}\step\hstep\object{B}\cr
	\vvbox{\hhbox{\id\Step\id\step\cd\step\id}
	       \hbox{\id\Step\hx\step\id\step\id}
	       \hhbox{\id\Step\id\step\ru\step\id}
	       \hbox{\cu\step\cu}
	       \hbox{\cd\step\cd}
	       \hhbox{\id\Step\id\step\rd\step\id}
	       \hbox{\id\Step\hx\step\id\step\id}
	       \hhbox{\id\Step\id\step\cu\step\id}}\cr
       \object{A}\Step\object{B}\step\hstep\object{A}\step\hstep\object{B}\cr}
\quad ={}\quad
\vvbox{\hhbox{\hstep\id\Step\step\id\step\cd\step\id}
       \hbox{\hstep\id\Step\step\hx\step\id\step\id}
       \hhbox{\hstep\id\Step\hstep\dd\step\ru\step\id}
       \hhbox{\hstep\id\Step\dd\step\cd\step\cd}
       \hhbox{\cd\step\cd\step\id\step\id\step\hrd\hstep\id}
       \hbox{\id\step\hx\step\id\step\id\step\hx\hstep\id\hstep\id}
       \hhbox{\cu\step\cu\step\id\step\id\step\hru\hstep\id}
       \hhbox{\hstep\id\Step\d\step\cu\step\cu}
       \hhbox{\hstep\id\Step\hstep\d\step\rd\step\id}
       \hbox{\hstep\id\Step\step\hx\step\id\step\id}
       \hhbox{\hstep\id\Step\step\id\step\cu\step\id}}
\quad =
\end{displaymath}
\begin{quotation}
The second equality follow from bialgebra axiom for $A$ and $B$.
The third use that $B$ is A-module coalgebra and comodule algebra.
The 4th use associativity and coherence.
The 5th is crossed module axiom.
The 6th follow from bialgebra axiom for $A$.
\end{quotation}
\caption{Bialgebra axiom for $A\ltimes B$ (part 1).}
\label{proofbialgebra1}
\end{figure}

\newfigure
\begin{figure}
\begin{displaymath}
 =\quad
\vvbox{\hbox{\hstep\id\Step\hstep\id\step\cd\Step\id}
       \hbox{\hstep\id\Step\hstep\hx\Step\id\step\cd}
       \hhbox{\cd\step\hstep\dd\hstep\cd\step\cd\hstep\id\Step\id}
       \hbox{\id\step\id\step\hcd\hstep\id\step\hx\step\id\hstep\rd\step\id}
       \hhbox{\id\step\id\step\id\step\id\hstep\ru\step\ru\dd\step\id\step\id}
       \hbox{\id\step\hx\step\id\hstep\id\Step\hx\step\hstep\id\step\id}
       \hhbox{\id\step\id\step\id\step\id\hstep\rd\step\rd\d\step\id\step\id}
       \hbox{\id\step\id\step\hcu\hstep\id\step\hx\step\id\hstep\ru\step\id}
       \hhbox{\cu\step\hstep\d\hstep\cu\step\cu\hstep\id\Step\id}
       \hbox{\hstep\id\Step\hstep\hx\Step\id\step\cu}
       \hbox{\hstep\id\Step\hstep\id\step\cu\Step\id}}
\quad =\quad
\vvbox{\hbox{\hcd\step\cd\step\hcd\Step\hstep\cd}
       \hbox{\id\step\id\step\id\Step\hx\step\d\step\hstep\id\Step\id}
       \hbox{\id\step\id\step\id\step\cd\d\step\id\step\hstep\id\Step\id}
       \hbox{\id\step\id\step\hx\Step\id\step\id\step\id\step\hstep\rd\step\id}
       \hhbox{\id\step\id\step\id\step\d\step\cd\hstep\id\step\id\step\dd\step
                                                                 \id\step\id}
       \hbox{\id\step\id\step\id\step\hstep\hx\step\id\hstep\id\step\hx\step
							     \hcd\hstep\id}
       \hbox{\id\step\hx\step\hstep\id\step\k\hstep\hx\step\hx\step\id\hstep
                                                                          \id}
       \hbox{\id\step\id\step\id\step\hstep\hx\step\id\hstep\id\step\hx\step
							     \hcu\hstep\id}
       \hhbox{\id\step\id\step\id\step\dd\step\cu\hstep\id\step\id\step\d\step
                                                                \id\step\id}
       \hbox{\id\step\id\step\hx\Step\id\step\id\step\id\step\hstep\ru\step\id}
       \hbox{\id\step\id\step\id\step\cu\dd\step\id\step\hstep\id\Step\id}
       \hbox{\id\step\id\step\id\Step\hx\step\dd\step\hstep\id\Step\id}
       \hbox{\hcu\step\cu\step\hcu\Step\hstep\cu}}
\quad =
\end{displaymath}
\caption{Bialgebra axiom for $A\ltimes B$ (part 2).}
\label{proofbialgebra2}
\end{figure}

\newfigure
\begin{figure}
\begin{displaymath}
 =\quad
\vvbox{\hbox{\hcd\step\cd\step\hstep\hcd\step\cd}
       \hhbox{\id\step\id\step\rd\step\id\step\hstep\id\step\id\step\rd\step
									 \id}
       \hbox{\id\step\hx\step\id\step\id\step\hstep\id\step\hx\step\d\d}
       \hhbox{\id\step\id\step\cu\step\d\step\id\step\id\hstep\cd\step\cd
								 \hstep\id}
       \hbox{\id\step\id\step\hstep\d\step\hx\step\id\hstep\id\step\id\step\id
							\step\id\hstep\id}
       \hbox{\id\step\id\Step\hstep\hx\step\hx\hstep\id\step\hx\step\id
							       \hstep\id}
       \hbox{\id\step\id\step\hstep\dd\step\hx\step\id\hstep\id\step\id\step
                                                    \id\step\id\hstep\id}
       \hhbox{\id\step\id\step\cd\step\dd\step\id\step\id\hstep\cu\step\cu
								 \hstep\id}
       \hbox{\id\step\hx\step\id\step\id\step\hstep\id\step\hx\step\dd\dd}
       \hhbox{\id\step\id\step\ru\step\id\step\hstep\id\step\id\step\ru\step
									 \id}
       \hbox{\hcu\step\cu\step\hstep\hcu\step\cu}}
\quad ={}\quad
\vvbox{\hbox{\hcd\step\cd\step\hstep\hcd\step\cd}
       \hhbox{\id\step\id\step\rd\step\id\step\hstep\id\step\id\step\rd\step
                                                                        \id}
       \hbox{\id\step\hx\step\id\step\id\step\hstep\id\step\hx\step\id\step\id}
       \hhbox{\id\step\id\step\cu\step\d\step\id\step\id\step\cu\step\id}
       \hbox{\id\step\id\step\hstep\d\step\hx\step\id\step\hstep\id\step
                                                                  \hstep\id}
       \hbox{\id\step\id\Step\hstep\hx\step\hx\step\hstep\id\step\hstep\id}
       \hbox{\id\step\id\step\hstep\dd\step\hx\step\id\step\hstep\id\step
                                                                  \hstep\id}
       \hhbox{\id\step\id\step\cd\step\dd\step\id\step\id\step\cd\step\id}
       \hbox{\id\step\hx\step\id\step\id\step\hstep\id\step\hx\step\id\step\id}
       \hhbox{\id\step\id\step\ru\step\id\step\hstep\id\step\id\step\ru\step
                                                                        \id}
       \hbox{\hcu\step\cu\step\hstep\hcu\step\cu}}
\quad =:\quad
\matrix{\step\object{A\ltimes B}\Step\step\object{A\ltimes B}\step\cr
        \vvbox{\hbox{\cd\step\cd}
               \hbox{\id\Step\hx\Step\id}
               \hbox{\cu\step\cu}}\cr
       \step\object{A\ltimes B}\Step\step\object{A\ltimes B}\step\cr}
\end{displaymath}
\caption{Bialgebra axiom for $A\ltimes B$ (part 3).}
\label{proofbialgebra3}
\end{figure}

\newfigure
\thispagestyle{empty}
\addtocounter{page}{-1}
\begin{figure}
\begin{displaymath}
\matrix{\object{X}\hstep\step\object{A\!\ltimes\! B}\step\cr
        \vvbox{\hhbox{\id\step\cd}
               \hbox{\hx\step\id}
               \hhbox{\id\step\ru}
               \hbox{\id\step\rd}
               \hbox{\hx\step\id}
               \hhbox{\id\step\cu}}\cr
       \object{X}\hstep\step\object{A\!\ltimes\! B}\step}
\quad :=\quad
\matrix{\object{X}\step\hstep\object{A}\Step\hstep\object{B}\step\cr
	\vvbox{\hbox{\id\step\hcd\step\cd}
	       \hhbox{\id\step\id\step\id\step\rd\step\id}
	       \hbox{\hx\step\hx\step\id\step\id}
	       \hbox{\id\step\hx\step\hcu\step\id}
	       \hhbox{\id\step\id\step\d\step\id\step\hstep\id}
	       \hhbox{\id\step\id\step\hstep\ru\step\dd}
	       \hhbox{\id\step\id\step\hstep\Ru}
	       \hbox{\id\step\id\step\hstep\Rd}
	       \hhbox{\id\step\id\step\hstep\rd\step\d}
	       \hhbox{\id\step\id\step\dd\step\id\step\hstep\id}
	       \hbox{\id\step\hx\step\hcd\step\id}
	       \hbox{\hx\step\hx\step\id\step\id}
	       \hhbox{\id\step\id\step\id\step\ru\step\id}
	       \hbox{\id\step\hcu\step\cu}}\cr
        \object{X}\step\hstep\object{A}\Step\hstep\object{B}\step}
\quad ={}\quad
\vvbox{\hhbox{\id\step\cd\step\hstep\id}
       \hbox{\hx\step\id\step\hstep\id}
       \hbox{\id\step\ru\hstep\cd}
       \hhbox{\id\step\d\step\rd\step\id}
       \hbox{\id\step\hstep\hx\step\id\step\id}
       \hhbox{\id\step\hstep\id\step\ru\step\id}
       \hhbox{\id\step\hstep\id\step\Ru}
       \hbox{\id\step\hstep\id\step\Rd}
       \hhbox{\id\step\hstep\id\step\rd\step\id}
       \hbox{\id\step\hstep\hx\step\id\step\id}
       \hhbox{\id\step\dd\step\ru\step\id}
       \hbox{\id\step\rd\hstep\cu}
       \hbox{\hx\step\id\step\hstep\id}
       \hhbox{\id\step\cu\step\hstep\id}}
\quad ={}\quad
\vvbox{\hhbox{\id\step\cd\Step\id}
       \hbox{\hx\step\id\step\cd}
       \hbox{\id\step\ru\step\rd\step\id}
       \hhbox{\id\step\rd\step\id\step\id\step\id}
       \hbox{\id\step\id\step\hx\step\id\step\id}
       \hhbox{\id\step\ru\step\id\step\id\step\id}
       \hbox{\id\step\rd\step\ru\step\id}
       \hbox{\hx\step\id\step\cu}
       \hhbox{\id\step\cu\Step\id}}
\quad ={}
\end{displaymath}
\begin{quotation}
The second equality use that $X$ is $A$-module and comodule.
The third is crossed module axiom over $B$.
In the 4th we use that $B$-action is $A$-module and
$B$-coaction is $A$-comodule morphisms.
The 5th follow from crossed module axiom over $A$, associativity and coherence.
The 6th use that $A$ is bialgebra.
\end{quotation}
\caption{Crossed module axiom over crossproduct $A\ltimes B$ (part 1).}
\label{proofovercross1}
\end{figure}

\newfigure
\begin{figure}
\begin{displaymath}
 =\enspace
\vvbox{\hbox{\Rd\step\hcd\step\id}
      \hbox{\id\Step\hx\step\id\step\id}
      \hhbox{\id\step\hstep\dd\step\ru\step\id}
      \hbox{\id\step\hcd\step\id\step\cd}
      \hbox{\hx\step\id\step\id\step\rd\step\id}
      \hbox{\id\step\k\step\hx\step\id\step\id}
      \hbox{\hx\step\id\step\id\step\ru\step\id}
      \hbox{\id\step\hcu\step\id\step\cu}
      \hhbox{\id\step\hstep\d\step\rd\step\id}
      \hbox{\id\Step\hx\step\id\step\id}
      \hbox{\Ru\step\hcu\step\id}}
\matrix{
=\enspace
\vvbox{\hbox{\Rd\Step\hcd\step\hstep\cd}
       \hhbox{\id\Step\d\step\dd\step\d\step\rd\step\d}
       \hbox{\id\Step\hstep\hx\Step\hx\step\d\hstep\d}
       \hhbox{\rd\step\dd\step\d\step\dd\hstep\cd\step\cd\step\id}
       \hbox{\id\step\hx\Step\hx\step\id\step\hx\step\id\step\id}
       \hhbox{\ru\step\d\step\dd\step\d\hstep\cu\step\cu\step\id}
       \hbox{\id\Step\hstep\hx\Step\hx\step\dd\hstep\dd}
       \hhbox{\id\Step\dd\step\d\step\dd\step\ru\step\dd}
       \hbox{\Ru\Step\hcu\step\hstep\cu}}
\enspace =\enspace
\vvbox{\hbox{\Rd\Step\hcd\step\hstep\cd}
       \hhbox{\id\Step\d\step\dd\step\d\step\rd\step\id}
       \hbox{\id\Step\hstep\hx\Step\hx\step\id\step\id}
       \hhbox{\rd\step\dd\step\d\step\dd\step\cu\step\id}
       \hbox{\id\step\hx\Step\hx\Step\id\hstep\step\id}
       \hhbox{\ru\step\d\step\dd\step\d\step\cd\step\id}
       \hbox{\id\Step\hstep\hx\Step\hx\step\id\step\id}
       \hhbox{\id\Step\dd\step\d\step\dd\step\ru\step\id}
       \hbox{\Ru\Step\hcu\step\hstep\cu}}
\enspace =:\cr
\begin{picture}(1,1)\end{picture}\cr
\qquad\quad\object{X}\Step\step\object{A\ltimes B}\step\cr
\quad =:\quad
        \vvbox{\hbox{\rd\step\cd}
               \hbox{\id\step\hx\Step\id}
	       \hbox{\ru\step\cu}}\cr
\qquad\quad\object{X}\Step\step\object{A\ltimes B}\step}
\end{displaymath}
\caption{Crossed module axiom over crossproduct $A\ltimes B$ (part 2).}
\label{proofovercross2}
\end{figure}

\end{document}